\documentclass[preprint,12pt]{aastex}

\newcommand{\teff}{T$_{eff}$}
\newcommand{\ki}{\ion{K}{1}}

\newcommand{\meth}{CH$_4$}
\newcommand{\wat}{H$_2$O}

\slugcomment{Submitted to ApJ 31 Aug 2006; accepted 15 Nov 2006}

\shorttitle{L/T Binaries}
\shortauthors{Burgasser}

\begin{document}

\title{Binaries and the L Dwarf/T Dwarf Transition}

\author{Adam J.\ Burgasser\altaffilmark{1}}
\affil{Massachusetts Institute of Technology, Kavli Institute for Astrophysics and Space Research,
Building 37, Room 664B, 77 Massachusetts Avenue, Cambridge, MA 02139; ajb@mit.edu}

\altaffiltext{1}{Visiting Astronomer at the Infrared Telescope Facility, which is operated by
the University of Hawaii under Cooperative Agreement NCC 5-538 with the National Aeronautics
and Space Administration, Office of Space Science, Planetary Astronomy Program.}

\begin{abstract}
High-resolution imaging has revealed an unusually high
binary fraction amongst objects spanning the transition between the
L dwarf and T dwarf spectral classes.  In an attempt to reproduce and unravel the origins
of this apparent binary excess, I present a series of
Monte Carlo mass function and multiplicity simulations of 
field brown dwarfs in the vicinity of the Sun.  These simulations
are based on the solar metallicity brown dwarf evolutionary models
and incorporate empirical luminosity and absolute magnitude scales,
measured multiplicity statistics 
and observed spectral templates in the construction and classification
of composite binary spectra.
In addition to providing predictions on the number
and surface density distributions of L and T dwarfs  
for volume-limited and magnitude-limited
samples, these simulations 
successfully reproduce the observed binary fraction distribution
assuming an intrinsic (resolved) binary fraction of 11$_{-3}^{+6}$\%
(95\% confidence interval), consistent with prior determinations.
However, the true binary fraction may be as high as 40\% if, as suggested
by Liu et al., a significant fraction of L/T transition objects 
($\sim$66\%) are tightly-bound, unresolved multiples.
The simulations presented here demonstrate that the binary excess
amongst L/T transition objects
arises primarily from the flattening
of the luminosity scale over these spectral types and is not inherently
the result of selection effects incurred in current magnitude-limited 
imaging samples.  Indeed, the
existence of a binary excess can be seen as further 
evidence that brown dwarfs traverse the L/T transition rapidly,
possibly driven by a nonequilibrium submergence of photospheric condensates.
%Following the supposition of Liu et al.
%that as many as two-thirds of L7--T5 dwarfs with parallax measurements are 
%overluminous, unresolved binaries, these simulations infer 
%an intrinsic brown dwarf binary fraction as high as 40\%,
%suggesting that brown dwarf binaries may be more common 
%and more tightly bound than commonly believed.
\end{abstract}


\section{Introduction}

L dwarfs and T dwarfs are the two lowest-luminosity spectral classes of low
mass stars and brown dwarfs known \citep[and references therein]{kir05}, 
spanning effective temperatures 
700 $\lesssim$ {\teff} $\lesssim$ 2300~K and luminosities
-5.7 $\lesssim \log{L_{bol}/L_{\sun}} \lesssim$ -3.5
\citep{gol04,vrb04}.  Their photometric and spectral
properties are remarkably distinct.  
L dwarfs typically exhibit red near-infrared
colors ($J-K \approx 1.5-2.0$), largely due to the presence of
optically thick condensate dust in their photospheres.  L dwarf spectra
are characterized by weak bands of metal oxides (with the exception
of CO), strong bands of
metal hydrides (including FeH, CrH and {\wat}) and strong neutral metal lines, 
in particular alkali species.  Cooler T dwarfs, on the other hand,
appear to have little or no condensate dust in their photospheres, exhibit
prominent {\meth} and {\wat} bands and pressure-broadened 
H$_2$ absorption in their near-infrared spectra, and generally have
blue near-infrared colors ($J-K \approx 0-0.5$).
There are $\sim$450 L dwarfs and $\sim$100 T dwarfs currently 
known,\footnote{See \url{http://www.dwarfarchives.org}.}
most identified in wide-field imaging surveys such as the Two Micron
All Sky Survey \citep[hereafter 2MASS]{skr06},
the Deep Near-Infrared Survey of the Southern Sky 
\citep[hereafter DENIS]{epc97} and the Sloan Digital Sky Survey 
\citep[hereafter SDSS]{yor00}.

The disparate properties of L dwarfs and T dwarfs
have fostered attention on the transition between these two spectral classes,
a transition that has proven to be rather unusual.
It was recognized early on that the range of effective
temperatures spanning the L/T transition\footnote{The L/T transition
is defined here as spanning spectral types L8--T4; cf.~\citet{leg00}.} 
is small, perhaps 100-300~K
\citep{kir00,gol04,vrb04}.
This is surprising, given the dramatic changes in spectral morphology
that occur over this range.  Parallax measurements of field sources  
further revealed an unexpected brightening of 
$\sim$1~mag between the latest-type L dwarfs and earliest-type
T dwarfs in the $J$-band ($\lambda_c \sim 1.2~\micron$; 
\citet{dah02,tin03,vrb04}).  This is also
surprising, given the slow decline in bolometric luminosities
over this range. Other unexpected trends across the L/T transition
include the observed resurgence in gaseous FeH bands, a species that is expected to be 
depleted by the formation of condensates in the L dwarfs \citep{mecloud,mcl03}; 
a restrengthening of $J$-band {\ki} lines after weakening considerably 
in the latest-type L dwarfs \citep{me02a,mcl03,cus05}; 
and enhanced CO absorption at 4.7~$\micron$, 
possibly due to vertical upwelling \citep{nol97,opp98,leg02,gol04}.
The current generation of brown dwarf atmosphere models incorporating condensate clouds
have been unable to reproduce these
trends \citep{ack01,coo03,tsu03,bur06}, so
dynamical (nonequilibrium) processes such as
cloud fragmentation \citep{mecloud},
a change in condensate sedimentation efficiency \citep{kna04}
and a global collapse of the condensate cloud layer \citep{tsu05} have
been evoked as possible drivers of the L/T transition.
However, persistent empirical uncertainties, such as the unknown ages and 
surface gravities of field dwarfs \citep{tsu03} and the role of unresolved
binaries \citep{bur06} have obviated conclusive results.

Recent high resolution imaging studies 
have helped to clarify our empirical understanding of the L/T transition.
These studies
have identified binaries whose components straddle the transition
\citep{cru04,mcc04,megl337cd,me0423,mehst2,liu06,rei2252}, 
two of which ---
SDSS J102109.69-030420.1AB, a T1 + T5 pair (hereafter SDSS~1021-0304); and
SDSS J153417.05+161546.1AB, a T1.5 + T5.5 pair  (hereafter SDSS~1534+1615) --- have been 
particularly revealing.
In both systems, the later-type secondary is the brighter source
at $J$-band, indicating that the brightening trend observed
in the parallax data is truly an intrinsic feature
of the L/T transition \citep{mehst2,liu06}.  However, the
brightening between the binary components, of order $\sim$0.2~mag, 
is significantly less than previously inferred.
Both \citet{mehst2} and \citet{liu06} have hypothesized that
prior results have been biased by
unresolved multiple systems.  Indeed, \citet{mehst2} have
demonstrated that the binary fraction of late-type L and early-type T dwarfs
is as high as 40\%, twice that of earlier-type L and
later-type T dwarfs.  This is likely a lower limit given the probable
existence
of more tightly bound, unresolved systems.  
Indeed, \citet{liu06} hypothesize that most apparently single
early-type T dwarfs may have unresolved companions.
Both studies ascribe this binary excess
to the contamination of magnitude-limited samples by systems consisting of
earlier-type and later-type components, whose composite spectra
mimic an L/T transition object.  While \citet{mehst2} demonstrated that the
binary excess can be roughly reproduced assuming
a rapid evolution of brown dwarfs across the transition,
several additional factors --- including the uncertain cooling rate
and absolute magnitude scale of brown dwarfs across the transition,
the underlying mass ratio distribution and frequency of brown dwarf
binaries, and possible systematic effects arising in magnitude-limited
samples --- were not considered.  Hence, the
origin of the binary excess and its 
implications on the properties of the L/T transition remain unsettled questions.

In this article, I present Monte Carlo mass function and binary spectral synthesis 
simulations aimed at examining the
origin of the L/T transition binary excess and its dependence on various
underlying factors.  The empirical multiplicity
dataset is described in detail in $\S$~2.  In $\S$~3, the construction and implementation
of the simulations are described, including realizations of the underlying mass function, binary frequency, mass ratio distribution, empirical luminosity and absolute magnitude trends, and the classification of unresolved binaries.  Results are presented in $\S$~4, 
focusing on number density, surface density and binary fraction 
distributions for local volume-limited and
magnitude-limited samples, and their dependencies on various 
input parameters.  Results are discussed in $\S$~5, including
a breakdown of the origin of the binary excess and an examination
of the intrinsic brown dwarf binary fraction based on the possible presence of 
overluminous, unresolved sources
in current parallax samples. Conclusions are summarized in $\S$~6.

\section{The Observed Binary Fraction}

The empirical multiplicity sample used here is similar to
that constructed in \citet{mehst2}, and
is based on results from high resolution imaging 
studies using the {\em Hubble
Space Telescope} (HST) and ground-based adaptive optics instrumentation
\citep{mar99,rei01b,rei06,bou03,mehst,mehst2,clo03,giz03}.
These studies were chosen as they represent a relatively uniform set 
of observations in terms of angular resolution 
(limits of 0$\farcs$05--0$\farcs$1)
and imaging sensitivity.  The selection criteria for sources imaged 
in these programs vary, but nearly all are composed of
objects identified in color-selected, magnitude-limited
surveys based primarily on 2MASS, DENIS and SDSS data.  It is therefore
assumed that, to first order, these data collectively 
constitute a magnitude-limited sample.
Of course, there is no single magnitude limit nor filter band 
to which that limit corresponds that spans the entire sample.
As I will show in $\S$~4.2.3, this detail appears to be relatively
unimportant in the resulting binary fraction distribution.

For all of the sources observed in these imaging studies,
spectral types were determined from published
optical classifications for L0--L8 dwarfs (on the \citet{kir99} scheme) 
and near-infrared data for L9--T8 dwarfs (on the \citet{geb02} and \citet{meclass} schemes).  This division in classification stems 
partly from the wavelength regimes over which the L and T
dwarf classes have been
defined, and specifically the current absence of a robust near-infrared
classification scheme for L dwarfs (see $\S$3.5.2).  Additionally,
few of the L dwarfs in these samples have reported near-infrared 
spectral data, while nearly all of them have been observed at
optical wavelengths.\footnote{Two exceptions to 
this are 2MASS J00452143+1634446 and DENIS J225210.7-173013,
both from the \citet{rei06} HST study, 
for which only near-infrared spectral types have been reported
(L3.5 and T0, respectively; \citet{wil03,rei06}).}
Conversely, while all of the T dwarfs in these samples have been
classified on the \citet{meclass} near-infrared scheme, few have
been observed in the optical \citep{me1237,me03opt,lie00}. 
In cases where the near-infrared spectrum of a source indicated
a spectral type of L9 or later, such as the L7.5/T0
SDSS J042348.57-041403.5 \citep[hereafter SDSS~0423-0414]{geb02,cru03}, the near-infrared
classification was retained.
Several sources in the \citet{bou03} HST sample required significant
revisions to their reported classifications
($>$3 spectral types) or were rejected given the absence of
any published spectral data.  Care was also taken to identify redundant sources
between the samples.  Finally, although the L2 dwarf Kelu~1 was unresolved
in HST observations by \citet{mar99}, more recent observations
by \citet{liu05} and \citet{gel06} have revealed this source to be 
binary, and is included as such in the sample.

Table~\ref{tab_bindata2} lists binary fractions for the full sample
in groups of integer subtypes, including 90\% and 95\% confidence
intervals based on the formalism of \citet{mehst}.
While the overall sample is fairly large (32 binaries in 162 systems),
significant uncertainties arise for individual subtypes due to 
small number statistics. The data were
therefore binned into four spectral type groups, 
L0--L3.5, L4--L7.5, L8--T3.5, and T4--T8, chosen so as to
minimize counting uncertainties while still sampling major
spectral subgroups. 
There is a clear increase in the fraction of binaries 
from early-type L to the L/T transition, peaking at
${\epsilon}_b^{obs}$ = 0.33$^{+0.19}_{-0.13}$ for the latter grouping.
Binaries are roughly twice as frequent amongst L/T transition objects
as compared to the 
earliest-type L dwarfs and latest-type T dwarfs.
However, even with the prescribed binning, uncertainties due
to small number statistics do not rule out a constant binary fraction
of 0.20--0.24 at the 90\% 
confidence level. Nevertheless, the simulations described below
will demonstrate that a binary excess amongst L/T transition objects
is an expected feature of the local brown dwarf population.

\section{Simulations}

\subsection{Motivation and Underlying Assumptions}

Both \citet{mehst2} and \citet{liu06} have proposed that the 
frequency of binaries
amongst L/T transition objects 
is a natural consequence of the properties of brown dwarfs in general.
In order to properly investigate
this hypothesis, it is therefore necessary to reproduce the empirical
sample as observed.
To do this, several basic assumptions were required.
First, as nearly all L and T dwarfs known reside
in the immediate vicinity of the
Sun, it was assumed that the distribution of these objects is spatially
isotropic.  It was also assumed that the entire population arises
from a single underlying mass function (MF) and age or birth rate distribution,
as described in $\S$~3.2.
Third, only single sources and binaries were modeled.  
Higher-order systems (triples, etc.) make up 
only 3\% of all very low mass multiples currently known
\citep{meppv}, and are therefore a negligible population for the purposes
of this study.
Fourth, the modeling of binaries assumes a fixed overall
binary fraction ($\epsilon_b$) and mass ratio distribution
($P(q)$, where $q \equiv$ M$_1$/M$_2$).  While there
is evidence that the binary fraction of stars decreases 
across several stellar classes
\citep{duq91,fis92,sha02,bou04}, no trend has
been established within the substellar regime.
Fifth, while simulated populations were initially constructed in
terms of space densities, they are re-sampled to reproduce
a magnitude-limited sample in order to investigate systematic
effects and make direct comparisons to the empirical data (see
$\S$~3.6).  Finally, it is assumed that all binaries generated by
the simulation are unresolved.  This constraint arises
from the observation that nearly all field L and T dwarf binaries 
identified to date have
apparent separations less than 1$\arcsec$ 
\citep{meppv}\footnote{See 
\url{http://paperclip.as.arizona.edu/$\sim$nsiegler/VLM\_binaries}.
Only one L dwarf binary wider than 1$\arcsec$ is currently known, 
the L1.5+L4.5 pair
2MASS J15200224-4422419AB \citep{ken06,me1520}. A handful of wide binaries 
with late-type M dwarf primaries and L or T dwarf secondaries are
also known \citep{bil05,me2200,bil06,mce06,rei06}.}, 
below the angular resolution limits of 2MASS, DENIS and SDSS.
Hence, a given magnitude-limited sample selected from these surveys
consistents entirely of singles and unresolved multiples.

\subsection{Generating a Brown Dwarf Population: Mass Function and Birthrate}

The first step of the simulation, generation of a population of 
low mass stars and brown dwarfs,
was done using the Monte Carlo code described in
\citet{memf}.\footnote{Comparable brown dwarf population
simulations
have also been presented in \citet{rei99,cha02,pir05,rya05,car06,dea06}; and
\citet{pin06}.
For an alternative Bayesian approach, see \citet{all05}.}  
Briefly, this code generates a large
set ($N = 10^6$)
of sources with randomly assigned masses and ages chosen
from the MF and age distributions. 
These sources are then assigned luminosities and
{\teff}s according to a set of evolutionary
models, in this case from \citet{bur01} and \citet{bar03}.  
The distribution of luminosities can
be compared to an observed luminosity function to constrain the underlying
MF and birth rate (e.g., Burgasser 2001, 2004; Allen et al.\ 2005).
Here, the derived luminosity functions provide a starting point for
examining the distribution of spectral types for single and multiple sources.

The input parameters for the simulation are listed in
Table~\ref{tab_parameters}.  Power-law forms of the MF,
$\Psi$(M) $\equiv dN/d{\rm M} \propto$ M$^{-\alpha}$, 
with $\alpha$ = 0.0, 0.5, 1.0 and 1.5 were examined; as well as 
the log-normal distribution of \citet[see also Miller \& Scalo 1979]{cha02}.  
A fixed mass range of 0.01 to 0.1~M$_{\sun}$
was assumed.  The lower mass limit is below the typical range of
masses that comprise field L and T dwarfs \citep{memf}, while
the upper limit is constrained 
by the maximum mass incorporated in the evolutionary models.
Note that the \citet{bur01} and \citet{bar03} models use
non-grey atmospheres and include condensate opacity as boundary
conditions but not in the emergent spectral energy distributions
(so-called ``COND'' models; \citet{all01}).  
Spectral models based on this assumption are largely
consistent with the properties of mid-type M and mid- and late-type T dwarfs
\citep{mar96,tsu96,tsu05};
but are generally inconsistent with those of late-type M 
and early-type L dwarfs in which condensate
opacity is important \citep{jon97,all01}. However, 
\citet{cha00} found $\lesssim$10\% difference 
in the evolution of 
luminosity and T$_{eff}$ between models that incorporate photospheric condensate 
dust and those that do not.  This is
a relatively small deviation given 
current empirical constraints on the brown dwarf
field luminosity function \citep{cru03} and binary fraction distribution 
(Table~\ref{tab_bindata2}). 
Hence, evolutionary tracks that incorporate condensate opacity 
were not included in the simulations.

The age distribution for the simulations assumes
a constant birth rate over the past 0.1-10~Gyr.
B04 have demonstrated that significant deviations in the resulting 
luminosity function
occur only for extreme forms of the birth rate, e.g., 
exponentially declining star formation rates
or short star-forming bursts.  These forms are generally inconsistent with
the observed continuity in the Galactic MF between low- and high-mass stars;
formation rates, kinematics and spatial distributions of planetary
nebulae, white dwarfs and \ion{H}{2} regions; nucleosynthesis yields; 
and the metallicity and activity distributions of G and K stars
\citep{mil79,sod91,boi99}.  Simulations with different 
minimum ages (0.1, 0.5 and 1~Gyr) showed no significant differences
in the output distributions (see $\S\S$~4.1.4 and 4.2.5).
The metallicities of the simulated sources were assumed
to be fixed at solar values, again a constraint of
the available evolutionary models.

Number densities from the simulations were normalized
to the field low-mass star (0.1--1.0 M$_{\sun}$) MF of \citet{rei99},
$\Psi$(M) = $0.35(\frac{\rm M}{0.1{\rm M}_{\sun}})^{-1.13}$ pc$^{-3}$ M$_{\sun}^{-1}$,
implying a number density of 0.0037 pc$^{-3}$
over the range 0.09--0.1 M$_{\sun}$.  
Because all of the distributions are scaled by this
factor, adjustment to refined measurements of the 
low-mass stellar space density can be readily made. 

\subsection{Assigning Spectral Types: The Luminosity Scale}

\begin{figure}
\epsscale{1.0}
\plotone{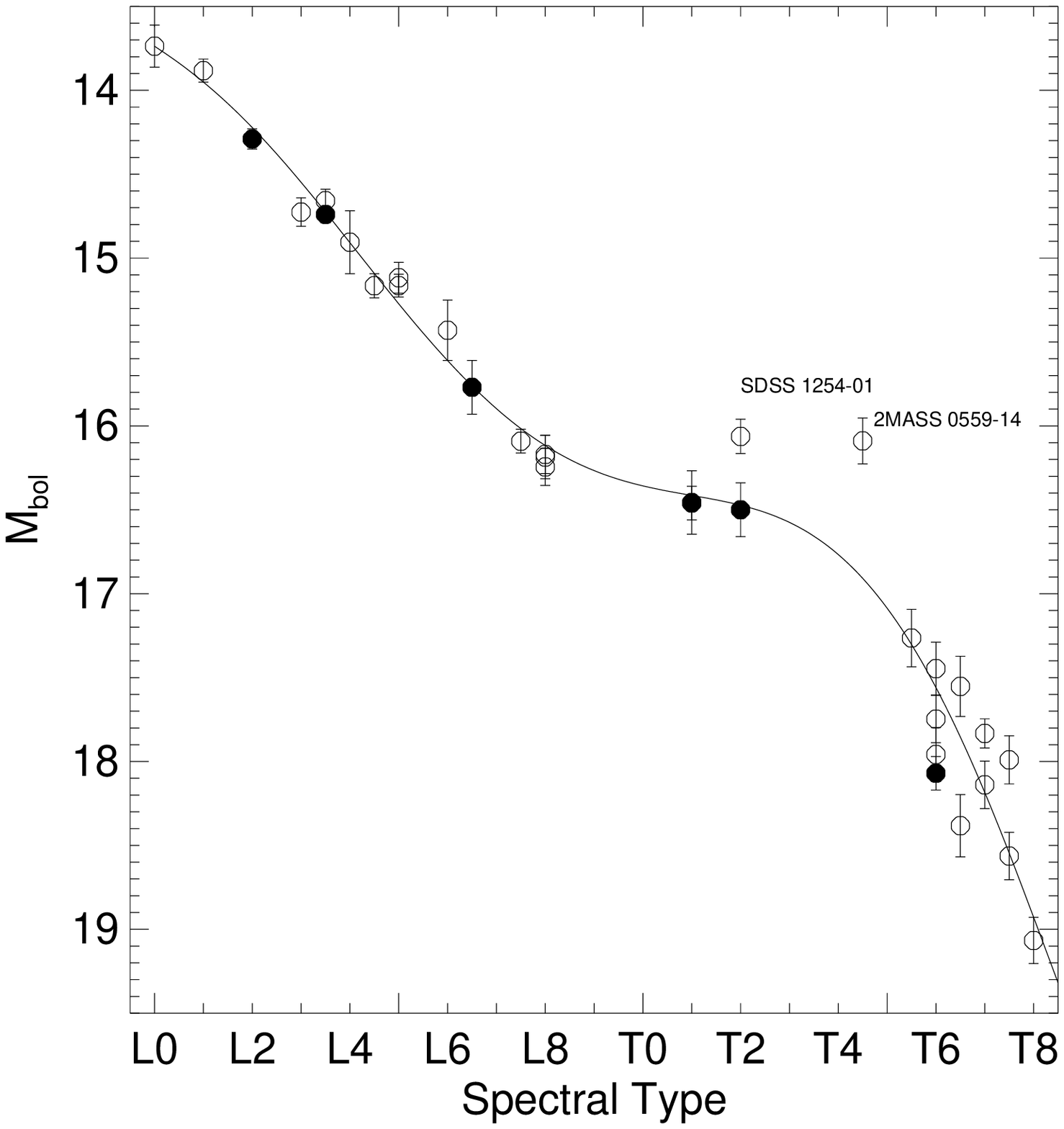}
\caption{Bolometic magnitude versus spectral type for 26 unresolved
field sources (open circles) and the six components of the binaries Kelu~1AB,
Epsilon Indi Bab and SDSS~0423-0414AB (filled circles).  Data are from
\citet{gol04,mcc04,liu05}; and \citet{mehst2}. A sixth-order polynomial fit
to the data, excluding the apparently overluminous sources SDSS 1254-0122 and
2MASS 0559-1404, is indicated
by the solid line (see Table~\ref{tab_fits}).  
The residual scatter in the fit is 0.22~mag.
\label{fig_mbol}}
\end{figure}

Once a sample of sources was generated and luminosities derived, 
spectral types were assigned using an
$M_{bol}$/spectral type relation based on empirical
measurements.  The data --- from \citet{gol04} for unresolved M6--T8 field
dwarfs, and from \citet{mcc04,liu05}; and \citet{mehst2} for
the components of the brown dwarf binaries 
Epsilon Indi Bab, Kelu 1AB and
SDSS J1021-0304AB, respectively ---  are shown in
Figure~\ref{fig_mbol}, and are restricted to sources
with $M_{bol}$ uncertainties $\leq$0.2~mag.   
As in the binary sample described in $\S$~2,
spectral types for the 32 sources in this plot
are based on the optical scheme
of \citet{kir99} for the L dwarfs (with the exception of the 
Kelu~1AB components; see
\citet{liu05}) and the near-infrared scheme of 
\citet{meclass} for the T dwarfs.  
Figure~\ref{fig_mbol}
also delineates a sixth-order polynomial fit to these data,
with coefficients listed in Table~\ref{tab_fits}.
This fit, which has a scatter of 0.22~mag, does not include measurements
for the overluminous T dwarfs SDSS J125453.90-012247.4 \citep[hereafter SDSS~1254-0122]{leg00}
and 2MASS J05591914-1404488 \citep[hereafter 2MASS~0559-1404]{me0559}.
Despite being unresolved in high-resolution imaging
observations \citep{mehst,mehst2}, these
sources are sufficiently overluminous for their spectral types
that they are likely to be
closely-separated binary systems \citep[see $\S$~5.2]{gol04,vrb04,liu06}.
Furthermore, were these
two sources included in the fit, the resulting 
$M_{bol}$/spectral type relation would be non-monotonic, obviating the ability 
to assign spectral types for a given
luminosity.  As it is, the $M_{bol}$ relation exhibits a 
remarkable flattening between types L8 and T4, decreasing by less than 0.5~mag
over this spectral type range.  This behavior is consistent with 
the small change in luminosity and {\teff} previously ascertained 
across the L/T transition.

An alternative method for assigning spectral types is to use the
simulated effective temperatures.  There have been several studies on
the {\teff}/spectral type scale for field L and T 
dwarfs\footnote{See \citet{bas00,kir00,me02a,dah02,leg02,schw02,gol04,nak04,vrb04}; and \citet{metgrav}.};
however, all of these
rely on an assumed radius (or distribution of radii) or spectral modeling,
and deviations between these methods have been noted
\citep{leg01,smi03}.  Because bolometric luminosity measurements
do not require any further theoretical assumptions, 
use of an $M_{bol}$/spectral type relation was favored.

\subsection{Incorporating Binary Systems}

Binary systems were incorporated by 
selecting $N_{bin} = {\epsilon}_bN$ of the simulated 
sources as primaries, and creating secondaries with the same
ages and masses assigned according to the mass ratio distribution.
Binary fractions spanning 0 $\leq {\epsilon}_b \leq$ 0.7
were examined, as well as two forms of $P(q)$: a flat distribution
and an exponential distribution, $P(q) \propto q^{\Gamma}$, where
$\Gamma$ = 4.  The latter form derives from statistical analyses of the
observed distribution of very low mass stellar and
brown dwarf binaries, which exhibits a strong
peak for equal-mass systems \citep{mehst2,rei06,all07}.
Luminosities, effective temperatures and spectral types were assigned
to the secondaries in the same manner as the primaries.
To facilitate computation, secondaries were forced to have masses
greater than the minimum mass sampled by the evolutionary models
(0.01~M$_{\sun}$).  Again,
for brown dwarfs with L and T dwarf luminosities, this constraint
had a negligible effect on the final distribution of secondaries.

\subsection{Composite Spectral Types}

\subsubsection{Generating Composite Spectra}

The binaries generated in these simulations
have combined light spectra determined by
the relative contributions of their primary and secondary
spectral types.  
As the absolute magnitude/spectral type
relations of L and T dwarfs
are distinctly non-linear \citep{dah02,tin03,vrb04}, 
and as the evolution of spectral features across 
this range is complex, 
the composite spectra were simulated explicity.
This was done by combining 
near-infrared spectral templates selected from a large library
(88 sources) of L and T dwarfs observed with 
the SpeX spectrograph \citep{ray03} in low-dispersion prism mode.  
Details on the acquisition and reduction of the SpeX prism data are described in \citet[see also Cushing et al.\ 2003 and Vacca, Cushing \& Rayner 2003]{metgrav}. 
The full library will be presented in a future
publication.  

\begin{figure}
\epsscale{1.0}
\plottwo{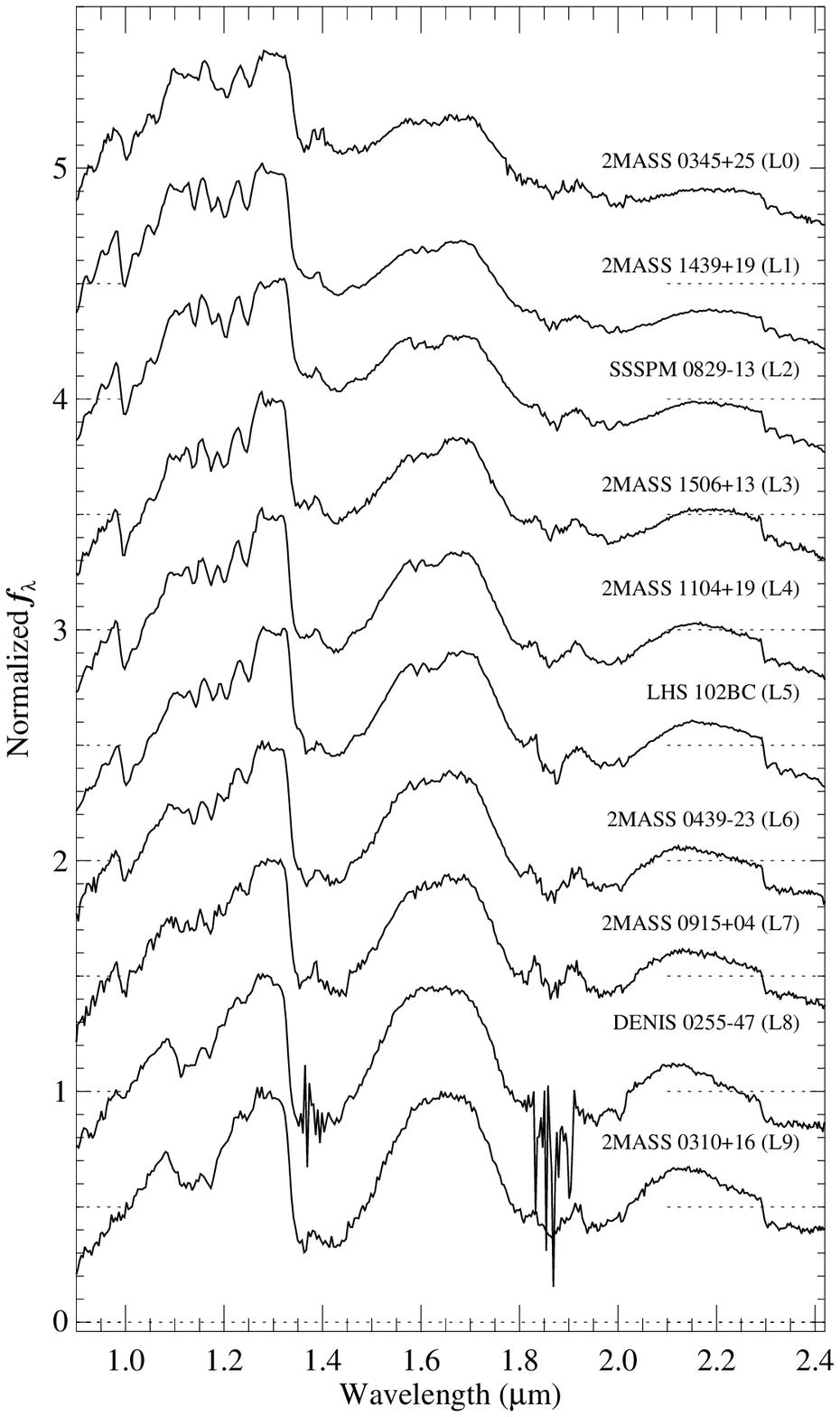}{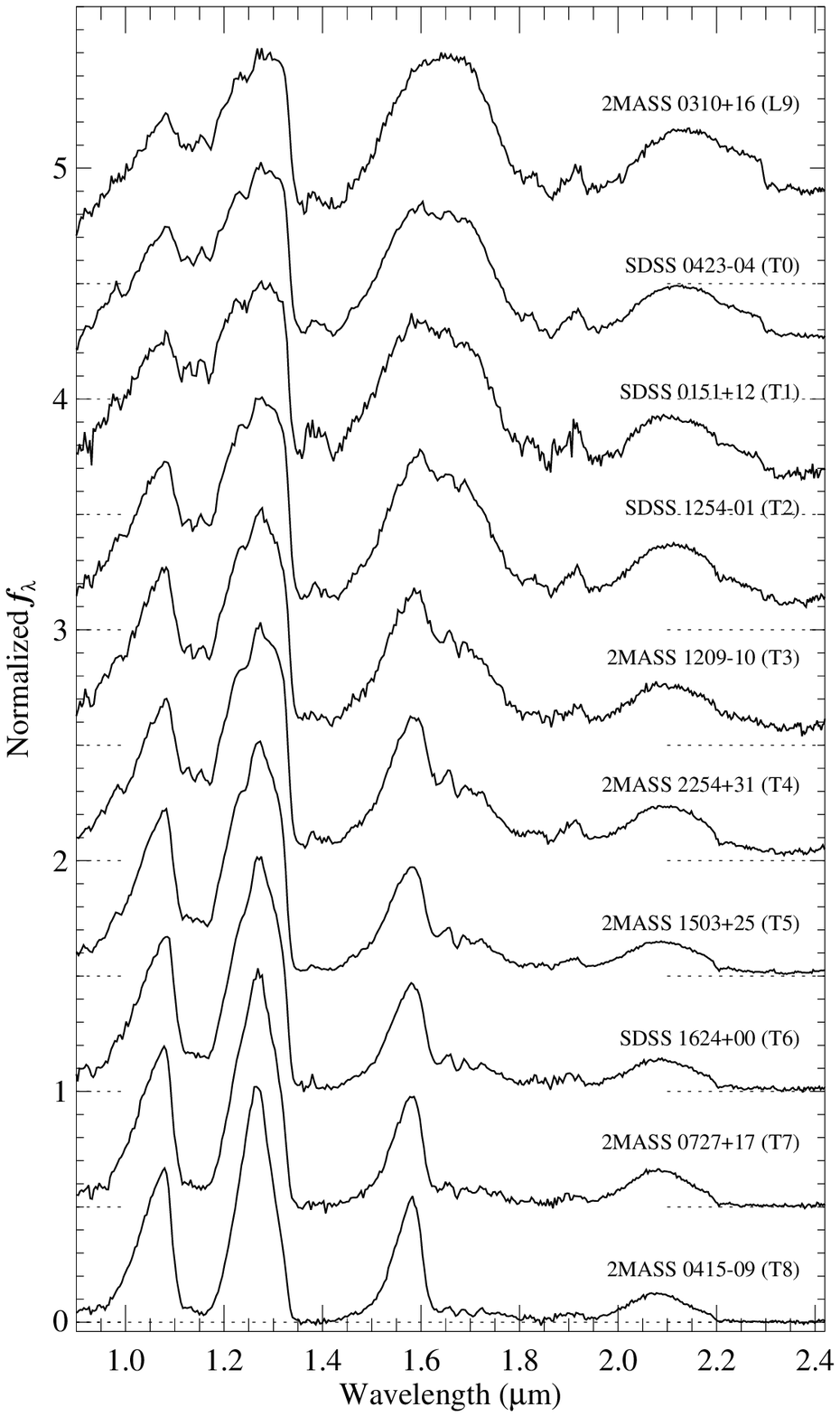}
\caption{SpeX spectral templates for L dwarfs (left) and T dwarfs (right).
Spectra are normalized in the 1.1-1.3~$\micron$ region and offset by
constants for clarity (dotted lines).  Names and spectral types are listed
(see Table~\ref{tab_standards}).
\label{fig_standards}}
\end{figure}

The spectral templates chosen are listed in 
Table~\ref{tab_standards} and their spectra shown 
in Figure~\ref{fig_standards}.
Selection of T dwarf templates was straightforward, as they
were simply chosen from the 
primary and secondary spectral standards 
from \citet{meclass}.
L dwarf near-infrared spectral standards, on the other hand,
have not yet been defined.  Furthermore,
substantial variance amongst the near-infrared 
spectra of L dwarfs with identical 
optical classifications has been noted \citep{geb02,mcl03,kna04,cus05,me1520}.
These variations likely arise from differences in the abundance 
of condensates in the photospheres of L dwarfs,
which is dependent on and/or augmented by surface gravity, 
metallicity and rotational effects \citep{ste03,kna04,bur06,chi06}.
As it is beyond the scope of this study to investigate these effects in
detail, L dwarf spectral
templates were conservatively chosen as sources with $J-K_s$
colors close to the mean of their spectral subtype \citep{kir00,vrb04}, 
and with no obvious spectral peculiarities or significant deviations
between published optical and near-infrared classifications.
Three of the L dwarf templates
are optical spectral standards in the \citet{kir99} classification scheme.
Three other L dwarf templates are binaries, but with components that have
similar magnitudes and photometric colors \citep{gol04,rei06,stu06}. 
Finally, an L9 standard, 2MASS J03105986+1648155 \citep{kir00}
was included as a tie between the L and T dwarf classes (cf.~\citet{geb02}).
As shown in Figure~\ref{fig_standards}, these spectra vary smoothly
in the evolution of the most dominant spectral features, including
{\wat}, CO and {\meth} absorption and overall spectral color.

\begin{figure}
\epsscale{1.0}
\plotone{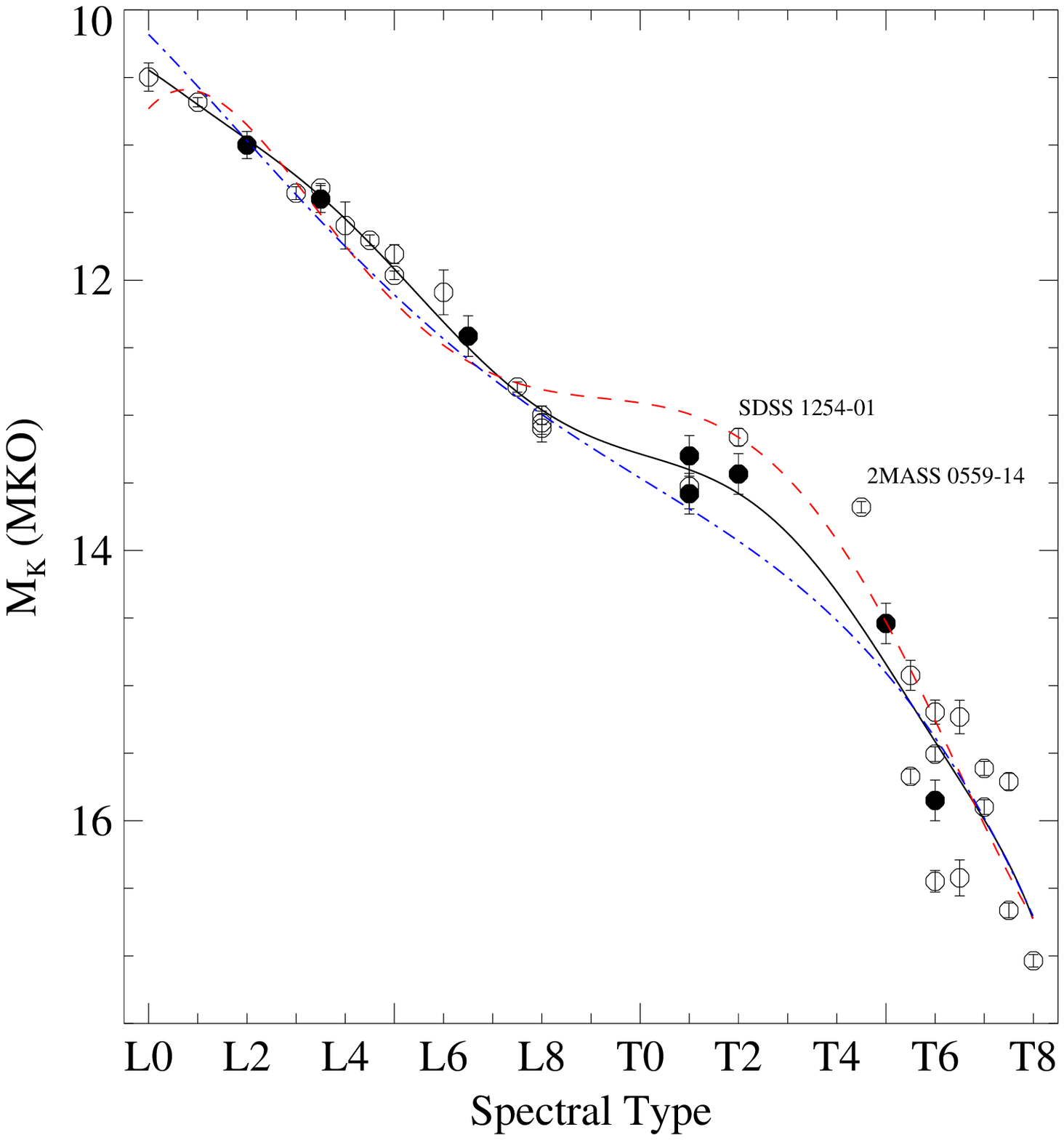}
\caption{Absolute MKO $K$-band magnitude versus spectral type
for 26 unresolved
field sources (open circles) and the eight components of the binaries Kelu~1AB,
Epsilon Indi Bab, SDSS~0423-0414AB and SDSS 1021-0304AB (filled circles).
Photometric data are from \citet{geb02,leg02,kna04,mcc04,liu05}; and
\citet{mehst2}.  Parallax data are from \citet{dah02,tin03}; and \citet{vrb04}.
A eighth-order polynomial fit to these data is indicated by the black line
(see Table~\ref{tab_fits}).
$M_K$/spectral type relations from \citet{liu06}, excluding known
binaries (red dashed line) and known and possible binaries (blue 
dot-dashed line) are also shown.
\label{fig_absk}}
\end{figure}

The spectral templates were then calibrated
to absolute fluxes according to an adopted absolute magnitude scale.
Three MKO\footnote{Mauna Kea Observatory system; \citet{sim02,tok02}.}
$M_K$/spectral type relations and two $M_J$/spectral type
relations were examined. 
The primary relation used in the simulations 
was derived by fitting an eighth-order polynomial to 
$M_K$ data\footnote{See \url{http://www.jach.hawaii.edu/$\sim$skl/LTdata.html}.} 
based on photometry \citep{geb02,leg02,kna04}
and parallax data \citep{dah02,tin03,vrb04} for 26 unresolved
field L and T dwarfs with combined uncertainties $\leq$0.2~mag;
and component absolute magnitudes for
the binaries Epsilon Indi Bab \citep{mcc04}, Kelu 1AB \citep{liu05}, 
SDSS J0423-0414AB and
SDSS J1021-0304AB \citep{mehst2}.
Figure~\ref{fig_absk} displays these data and the polynomial fit, 
which has a dispersion of 0.26~mag (coefficients are listed 
in Table~\ref{tab_fits}).  
Again, the overluminous sources 2MASS~0559-1404 and
SDSS~1254-0122 were excluded from the fit.
The other absolute magnitude/spectral type relations were taken
from \citet{liu06},
based on fits to the absolute magnitudes of field sources
excluding either known (``L06'') or known and possible binaries
(``L06p''; see $\S$~5.2).
Figure~\ref{fig_absk} compares the three $M_K$/spectral type relations.
The most significant
differences between these lie in the L8 to T5 range, 
over which the two relations of \citet{liu06} differ by
$\sim$1~mag.  The polynomial fit derived here is intermediate
between these two relations.
The $M_J$/spectral type relations considered here
are illustrated in Figure~4 of \citet{liu06}.

Once the template spectra were scaled to their 
respective absolute magnitudes
following standard techniques (cf.~\citet{cus05}),
composite spectra were created by adding the templates that
bracketed the assigned spectral types of the binary components,
weighted accordingly\footnote{For example, 
an L1.3+T4.7 composite spectrum was modeled
as 0.7$\times$L1 + 0.3$\times$L2 + 0.3$\times$T4 + 0.7$\times$T5.}.
Examples of composite spectra for various combinations of primary
and secondary spectral types are
illustrated in Figure~\ref{fig_hybrid}.

\subsubsection{Classifying the Composite Spectra}

\begin{figure}
\epsscale{1.0}
\plotone{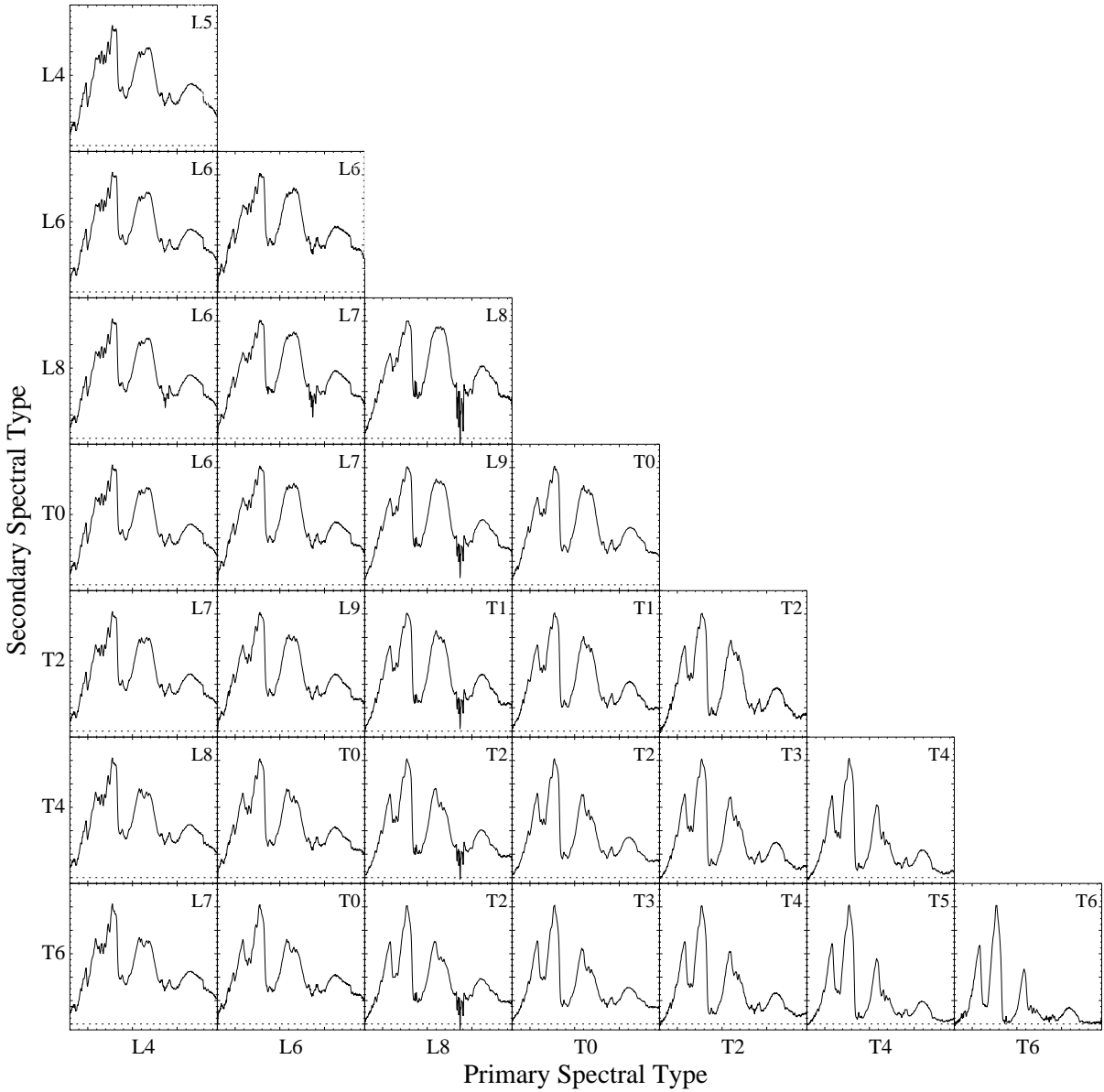}
\caption{Hybrid binary spectra for various combinations of primary (ordinate)
and secondary (abscissa) spectral type combinations.  All spectra
are normalized in the 1.1--1.3~$\micron$ band.  Composite spectral
types, based on spectral indices as described in $\S$~3.5.2, 
are indicated in the upper right corner of each box.
\label{fig_hybrid}}
\end{figure}

For binaries with large differences in their component spectral types,
composite spectra can differ substantially in their spectral
characteristics from the underlying standard sequence. 
Examples include the appearance
of {\meth} absorption at $H$-band and not $K$-band, as observed in the
L6: + T4: binary 2MASS J05185995-2828372AB 
\citep[cf.~Figure~\ref{fig_hybrid}]{cru04,mehst2}; or strong deviations
in individual {\wat} absorption bands.  As tens to
hundreds of thousands of composite spectra were generated 
in each simulation, an automated classification scheme was required
After experimenting with
various techniques, it was determined that classification by 
spectral indices provided the most efficient and accurate method.  

\begin{figure}
\epsscale{1.0}
\plotone{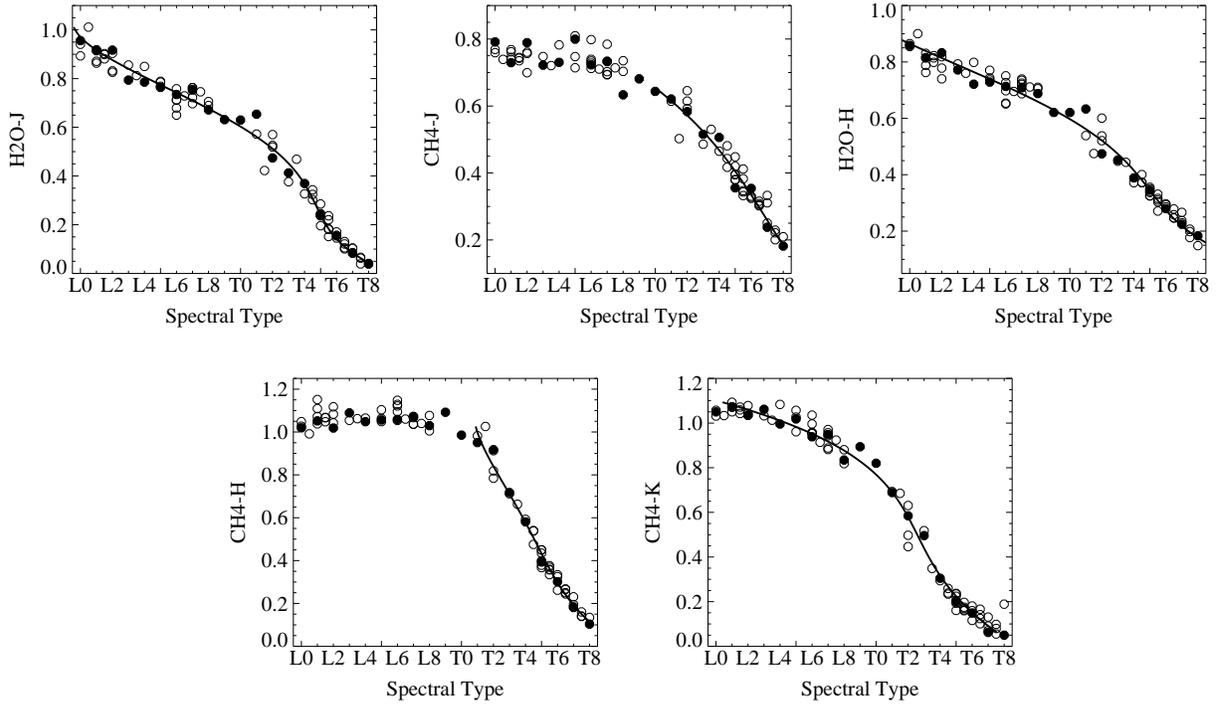}
\caption{Spectral indices as a function of spectral type for a large sample
of SpeX prism data.  Measurements made on the spectral templates shown in Figure~\ref{fig_standards} are indicated by filled circles, all others by open circles.  Spectral types are based on optical data for L0-L8 dwarfs
and near-infrared data for L9-T8 dwarfs.  Fourth-order polynomial fits for
each index over a prescribed range are indicated by solid lines
(see Table~\ref{tab_fits}).
\label{fig_indices}}
\end{figure}

For T dwarfs, classification by indices is straightforward,
as \citet{meclass} have defined five near-infrared indices ({\wat}-J, {\meth}-J, 
{\wat}-H, {\meth}-H and {\meth}-K) which track monotonically
with T spectral type.
For the L dwarfs, several near-infrared indices have been
defined and compared against optical types
\citep{rei01a,tes01,geb02,mcl03}, the most successful of
which are tied to the {\wat} absorption bands.
However, a robust classification scheme for L dwarfs in the near-infrared
is still lacking, which can lead to problems particularly
at the L/T transition \citep{chi06}.  I therefore
examined the behavior of the \citet{meclass} indices for both
L and T dwarf spectra from the SpeX prism library, as shown
in  Figure~\ref{fig_indices}.
Following the convention of the empirical binary sample,
spectral types
for L0-L8 dwarfs are based on optical data while those for L9-T8
dwarfs are based on near-infrared data.
There are strong correlations for three of these
indices --- the two {\wat} indices and {\meth}-K --- that are roughly
monotonic across full
spectral type range.
{\meth}-J and {\meth}-H are strongly correlated only
in the T dwarf regime.
Fourth-order polynomial fits to the indices were made where correlations
were strongest, and coefficients are listed in Table~\ref{tab_fits}. 

Spectral types for the composite spectra were determined from these spectal index relations, as follows.
First a spectral type was estimated from the {\wat}-J, {\wat}-H and
{\meth}-K indices.  
If these indices indicated
a T spectral type, or if {\meth} absorption is present at $H$-band
(indicated by {\meth}-H $<$ 0.97), subtypes based on the 
{\meth}-J and {\meth}-H indices were also computed.
The individual index spectral types were then 
averaged to derive an overall classification for the composite.
Figure~\ref{fig_hybrid} lists the 
classifications derived for the binary combinations shown there.

\begin{figure}
\epsscale{1.0}
\plotone{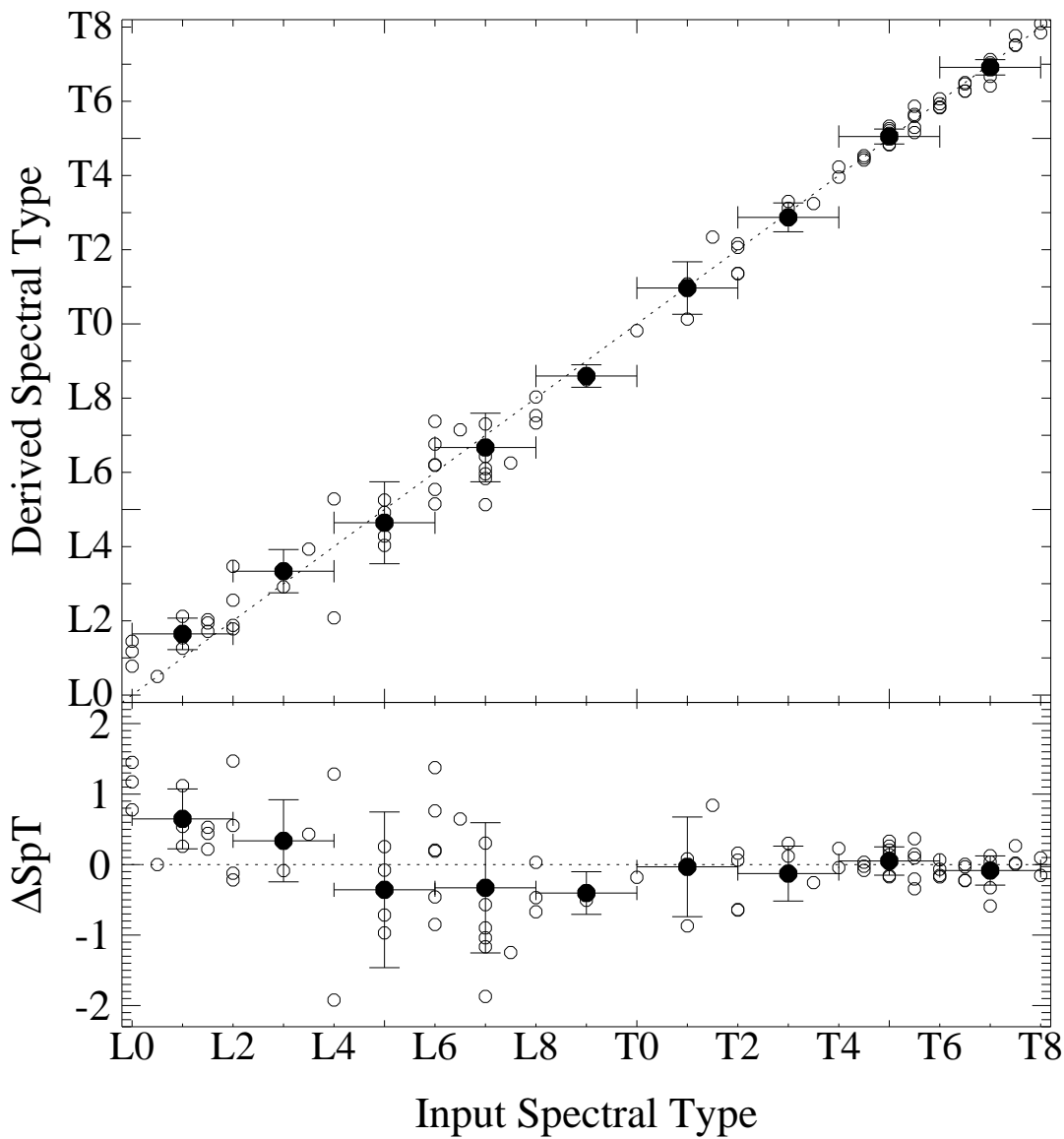}
\caption{Examination of the robustness of 
near-infrared classifications for L and T dwarfs using the
spectral index relations of Table~\ref{tab_fits}.  
Plotted in small open circles 
are the derived spectral types for 88 L0--T8 dwarfs with SpeX prism data
using the spectral index relations of Table~\ref{tab_fits}, compared to
their published optical (L0--L8) and near-infrared
(L9--T8) spectral types. Average deviations and 1$\sigma$ scatter
from perfect agreement (dotted line) in groupings of two spectral subtypes
are indicated by filled circles with error bars.  The typical
disagreement for the entire sample is 0.6~SpT, but is greater for 
L0--L8 dwarfs (0.9~SpT) than L9-T8 dwarfs (0.3~SpT).
\label{fig_classcomp}}
\end{figure}

To determine the robustness of this method, the L and T dwarf SpeX spectra 
used to derive the index relations were
reclassified by the above technique and compared to their original
classifications.  Figure~\ref{fig_classcomp} illustrates this
comparison and typical deviations.  Overall, spectral types derived from
the indices agree with published values
to within 0.6~subtypes, although the scatter is greater amongst L0-L8 dwarfs
(0.9~subtypes) than L9--T8 dwarfs (0.3~subtypes) or L8--T4 dwarfs
(0.4~subtypes).  These
deviations likely arise from the same photospheric condensate variations
that have inhibited the identification of near-infrared L dwarf spectral
standards \citep{ste03}.  
Nevertheless, Figure~\ref{fig_classcomp} indicates
an overall accuracy of 0.5-1.0 subtypes can be attained for both
L dwarfs and T dwarfs
using a common set of spectral indices.

\subsection{Constructing Magnitude-limited Samples}

Once the parameters for all single and composite
systems were derived, the simulated population
was resampled into integer spectral type bins to derive distributions in number
density and binary fraction.  These distributions were normalized 
according to the assumed
space density ($\S$~3.2), yielding a ``volume-limited'' sample.  The
resulting number densities were then converted to surface densities
for a magnitude-limited sample by computing effective volumes 
for each system,
\begin{equation}
V_{eff,i} = \frac{1}{3}10^{0.6(m_{lim} - M_i)+3}~{\rm pc}^3
\end{equation}
\citep{sch68},
where $m_{lim}$ is the adopted limiting magnitude and $M_i$ the 
absolute magnitude of the system (i.e., the absolute magnitude 
of a single source
or the combined light magnitude for a binary system).  For all simulations,
$m_{lim}$ = 16 was adopted as a proxy.
Surface densities to this magnitude limit
were then computed as the sum of effective volumes
over a given spectral type range,
\begin{equation}
\Sigma({\rm SpT)} = \sum_i^{i \in {\rm SpT}}2.424{\times}10^{-5}V_{eff,i}~{\rm deg}^{-2}
\end{equation}
where the numerical factor provides units of deg$^{-2}$.

\section{Results}

A total of 58 separate simulations were run to sample the various input parameters listed in Table~\ref{tab_parameters}.
From each simulation, volume-limited space densities, 
magnitude-limited surface densities,
and binary fraction distributions 
(for both volume-limited and magnitude-limited cases) 
as a function of primary, secondary and systemic 
(singles plus composite binaries)
spectral type were constructed.  
In the comparisons that follow, the baseline
simulation is based on the \citet{bar03} evolutionary models, 
a power-law mass function with $\alpha$ = 0.5, ${\epsilon}_b$ = 0.1,
the exponential mass ratio distribution, and the $M_K$/spectral
type relation derived in $\S$~3.5.1.

\subsection{Number and Surface Density Distributions}

While the focus of this study is the binary fraction distribution
of L and T dwarfs, several clues into its origin can be gleaned from
examining the number density (for the volume-limited case) and
surface density (for the magnitude-limited case) distributions 
and their dependencies on various underlying parameters. Results
from representative simulations are provided in Tables~\ref{tab_sptdist}
and~\ref{tab_sptdistmag}.

\subsubsection{Distribution of Single Sources and 
Dependencies on the Underlying Mass Function}

\begin{figure}
\epsscale{1.0}
\plottwo{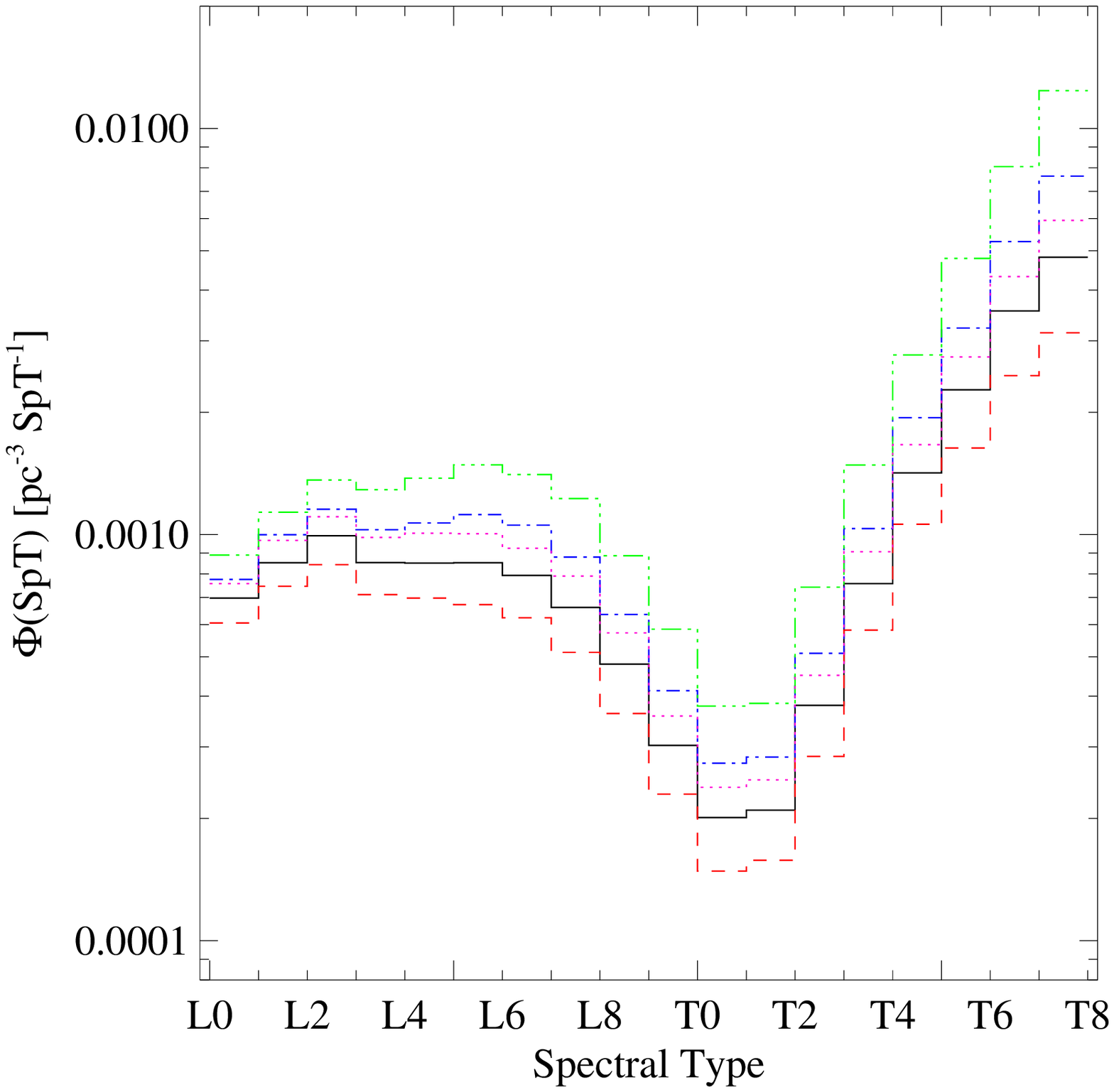}{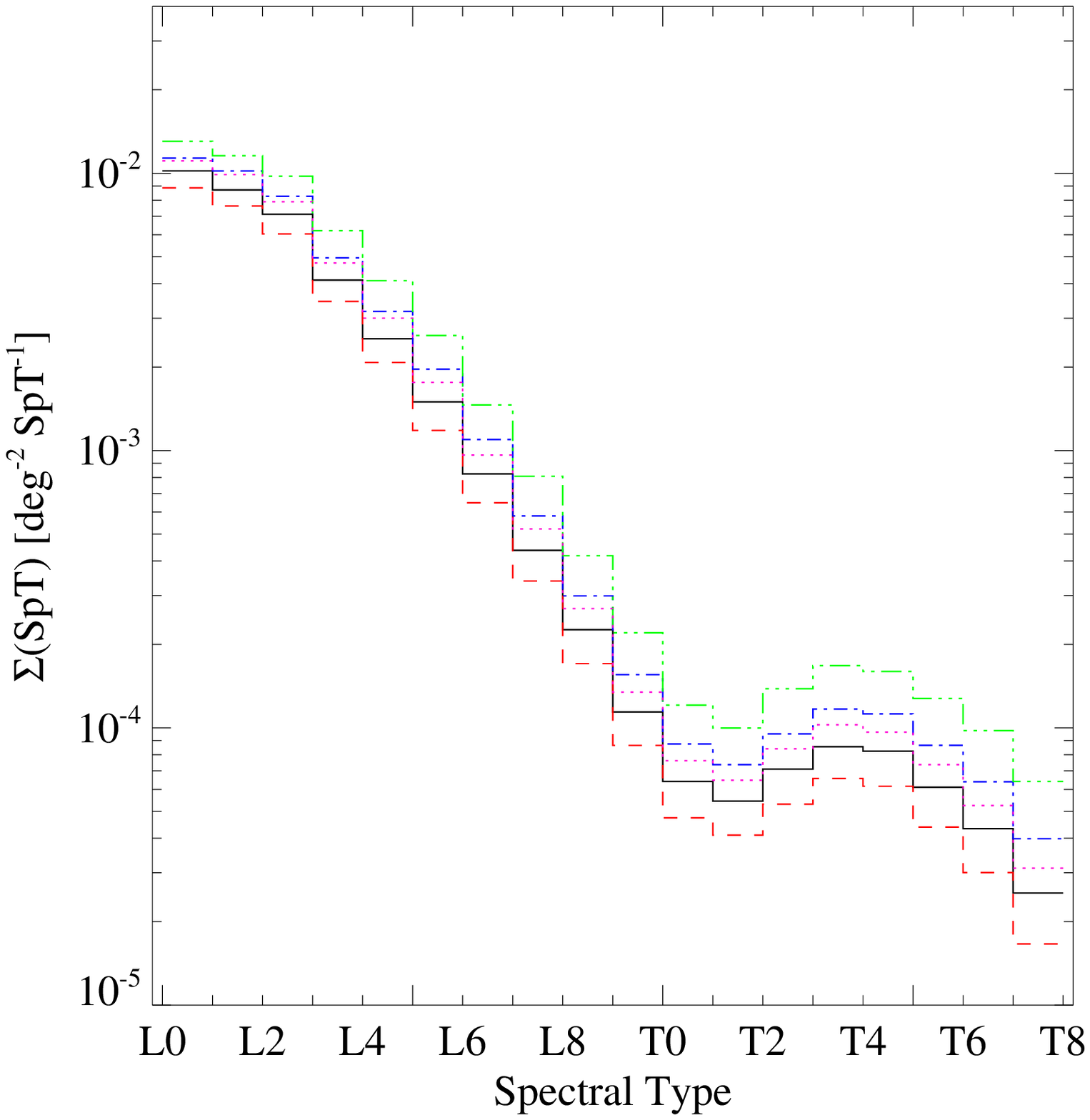}
\caption{Number ($\Phi$(SpT); left panel) and surface  
density ($\Sigma$(SpT); right panel) distributions for single 
sources based on simulations using power-law mass functions
with $\alpha$ = 0 (red dashed line), 0.5 (solid black line), 1.0 
(blue dot-dashed line) and 1.5 (green triple-dot-dashed line),
and the lognormal distribution of 
\citet[magenta dotted line]{cha02}. Number densities are
shown in units of pc$^{-3}$~SpT$^{-1}$ for a volume-limited sample, 
surface densities are shown in units of deg$^{-2}$~SpT$^{-1}$ 
for a magnitude-limited sample with $K \leq 16$.
\label{fig_sptvsmf}}
\end{figure}

Figure~\ref{fig_sptvsmf} plots the distribution of singles sources
(i.e., for ${\epsilon}_b$ = 0) as a function of spectral type
for volume-limited and magnitude-limited samples and the five
mass functions examined.  The overall shape of the volume-limited
distributions is similar for all of the mass functions, with 
a plateau\footnote{The slight rise 
in number densities from L0 to L2 is likely an artifact 
of the maximum mass (0.1~M$_{\sun}$) used in the simulations.}
amongst early- and mid-type L dwarfs, followed by
a decline in number density from mid-type L to a minimum around
T0--T2 (a factor of 4-5 decrease); then, a rapid rise 
toward the latest-type T dwarfs.  
The shape of these distributions
is similar in form to the simulated field brown
dwarf luminosity functions of \citet{memf} and \citet{all05},
and arises from the combination of long-lived
stellar-mass early-type L dwarfs (which accumulate
over the 10~Gyr of the simulation),
the rapid cooling rate of 
substellar-mass L dwarfs (reducing the number of late-type L dwarfs)
and the slow cooling rate of late-type T dwarfs (increasing their
numbers).  
The dip at T0--T2 is enhanced by the flattening in the
$M_{bol}$/spectral type relation over this range, which spreads the
luminosity function thinly at these spectral types.
In addition, mass functions with $\alpha > 0$
preferentially produce 
lower-mass brown dwarfs and hence contribute to the rise in number
density amongst the lowest-luminosity brown dwarfs.
Overall number densities increase for steeper power-law
mass functions, as expected, ranging over 15--60\% from types L0 to T8
for $\Delta\alpha$ = 0.5.  The log-normal mass function produces a
distribution intermediate between the $\alpha$ = 0.5 and 1.0 
distributions.
The number density of L0--L9 (T0--T8) dwarfs in these simulations range over
0.006--0.011~pc$^{-3}$ (0.009--0.03~pc$^{-3}$) for $\alpha$ = 0 to 1.5; i.e.,
there are 50--250\% more T dwarfs than L dwarfs in a given volume
for the mass functions examined.
The number density of L/T transition objects, L8--T4, range over
0.0018--0.004~pc$^{-3}$ for $\alpha$ = 0 to 1.5, and are thus comparatively 
rare.

In the magnitude-limited case, 
the number density decline from mid-type L to early-type T
combined with the decrease in absolute $K$-band brightness over this range
results in a steep decline in surface densities, by a factor of $\sim$120-200
from L0 to T1 depending on the underlying MF.  
Surface densities then increase by $\sim$60--70\%
over types T1--T4, driven
by both the increase in number densities and the flattening of the
$M_K$/spectral type relation
over this spectral type range.
Beyond T4, surface densities again drop as the
$M_K$/spectral type relation steepens.  The intrinsic
brightness of L dwarfs implies that there are 50--80 times more of these
objects than T dwarfs in a given
magnitude-limited sample, in stark contrast to 
the ratio of number densities.\footnote{The current tally of known L dwarfs and
T dwarfs, a ratio of $\sim$4.5:1, largely reflects preferential
efforts toward searches of later-type sources.  If both classes were investigated to similar 
($K$-band) magnitude limits, perhaps 5000-8000 L dwarfs would be known.} 
Again, steeper mass
functions result in more L and T dwarfs, by roughly the same factor
as the volume-limited case.

\subsubsection{Distributions of Primaries, Secondaries and Systems}

\begin{figure}
\epsscale{1.0}
\plottwo{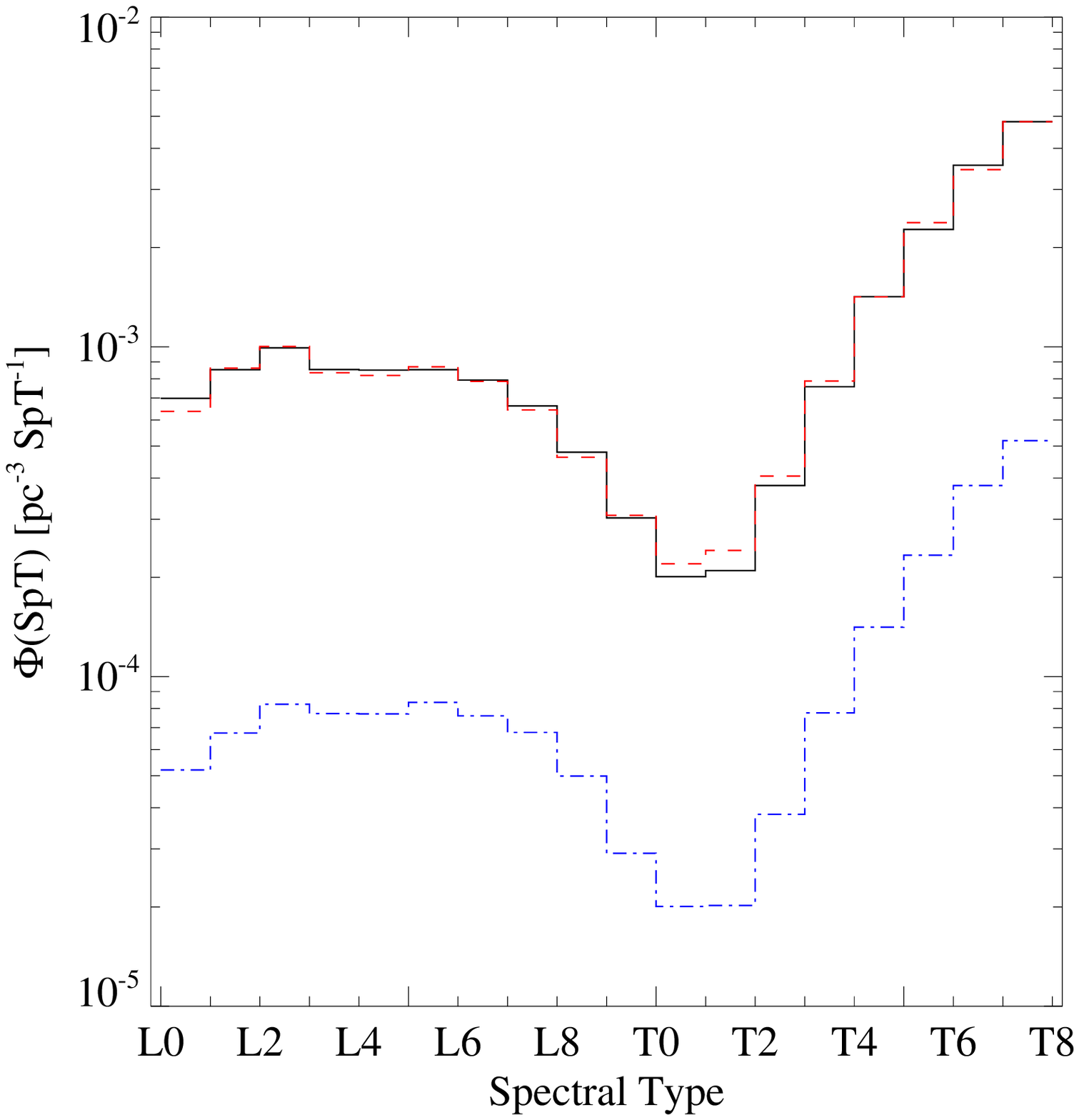}{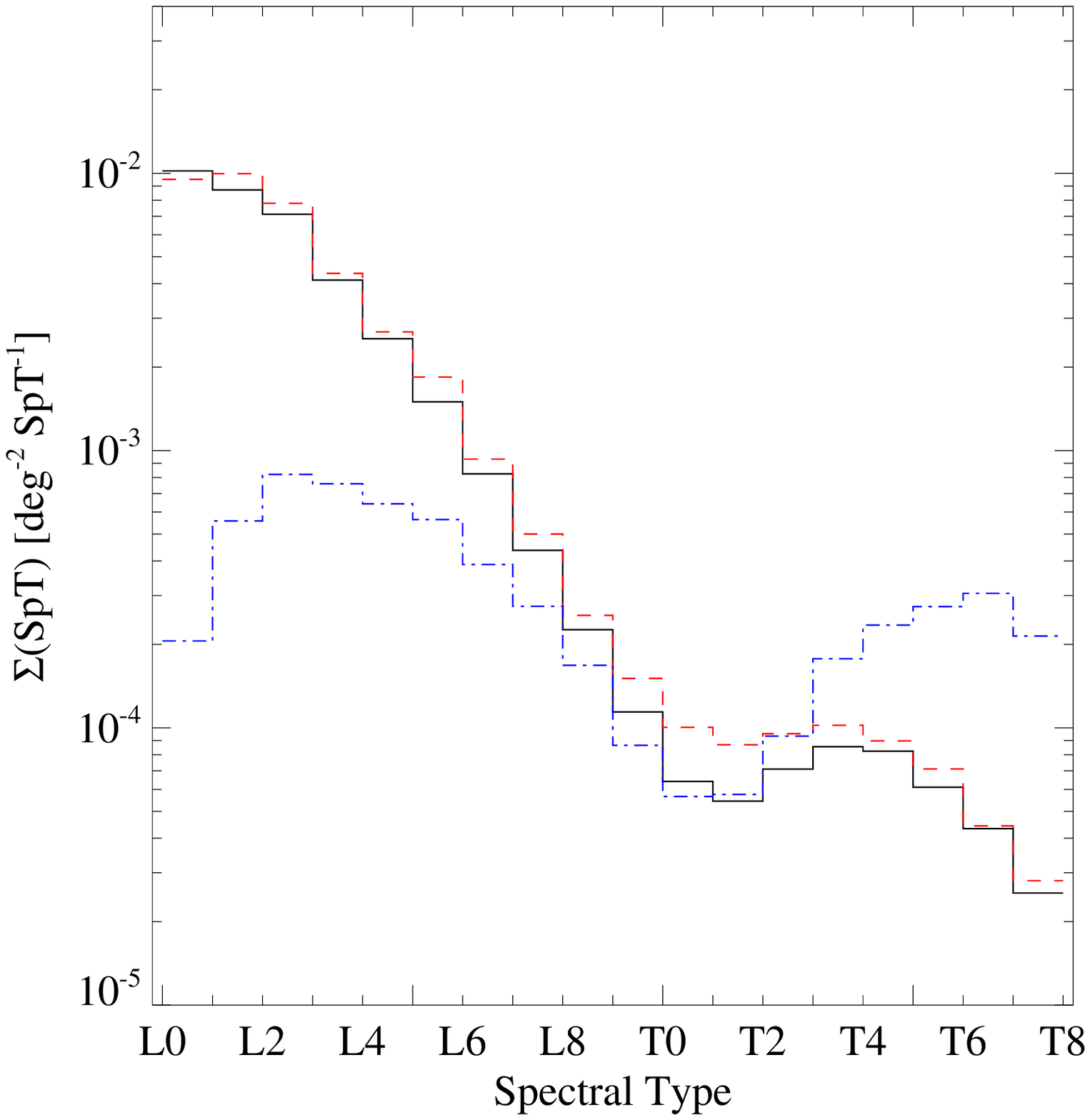}
\caption{Number (left panel) and surface density (right panel) 
distributions for primaries (solid black lines), secondaries 
(blue dot-dashed lines) and systems (red dashed lines)
for the baseline simulations. Surface densities are based on 
a magnitude-limited sample ($K \leq 16$).
\label{fig_sptvscomp}}
\end{figure}

Figure~\ref{fig_sptvscomp}
compares the spectral type 
distributions of primaries, secondaries and
systems (singles plus binaries) 
in both volume-limted and magnitude-limited cases.
The only clear difference seen between the primary and systemic
number density distributions in the volume-limited case is a 
10-15\% increase in the number of T0--T2 dwarf systems.
This deviation increases for higher binary fractions,
%reaching 80\% for ${\epsilon}_b$ = 0.5.
and while relatively small %deviation, we shall see this
nevertheless has a significant impact on the binary fraction distribution.
For the magnitude-limited
distribution, the contribution of binaries in the surface 
densities of T0--T2 dwarfs increases substantially,
by $\sim$60\% for ${\epsilon}_b$ = 0.1, implying that
40\% of these objects are binary.  This is consistent with the 
observed binary fraction for these spectral types
(Table~\ref{tab_bindata2}).
Increasing the underlying binary fraction raises the entire
surface density distribution in the magnitude-limited case, 
as the overluminous, unresolved binaries are sampled
to larger distances and greater volumes.  However, the largest
increase is consistently seen in the surface densities of T0--T2 dwarfs,
reaching a factor of 4 increase for ${\epsilon}_b$ = 0.5.

The shape of the number density distribution of 
secondaries in Figure~\ref{fig_sptvscomp} 
closely follows that of the primaries and composites, but
is scaled roughly by the underlying binary fraction.
There is a slight
increase in the relative numbers of T dwarf secondaries,
an artifact of the constraint that secondaries have equal or lower
masses than their primaries. 
Since M dwarf primaries are not modeled here, there is an artificial
deficit of L dwarf secondaries.  
This bias is greatly enhanced
in the magnitude-limited case.
Interestingly,
the number of T dwarf secondaries is substantially larger
than the number of primaries
in this case, by a factor of 3 or more for spectral
types T4 and later and ${\epsilon}_b$ = 0.1.
The vast majority of these secondaries are
low mass companions to L dwarfs, as their relative infrequency
(lying on the tail of the power-law mass ratio distribution)
is outweighed by the much larger number of brighter
L dwarf primaries in a magnitude-limited sample.
A significant fraction of these systems consistent of old, 
low-mass stellar (L dwarf) + brown dwarf (T dwarf)
pairs whose components straddle the hydrogen burning limit. 
These results indicate that, for a given
magnitude limited sample, many more T dwarfs may be found as companions
to higher mass brown dwarfs and/or stars than as isolated 
objects,\footnote{There is an indication of this trend 
when one considers that of all known T dwarfs with distance 
measurements within 8 pc, two are isolated field objects
(2MASS J04151954-0935066 and 2MASS J09373487+2931409; 
\citet{me02a,vrb04}),
while five are companions to nearby stars
(Epsilon Indi Bab; Gliese 229B, Gliese 570D and SCR 1845-6357B;
\citet{nak95,megl570d,sch03,mcc04,bil06}).  However, 
selection effects, including concentrated efforts 
to find low mass companions to nearby stars, may
be responsible.}
although uncovering these companions
may be nontrivial. 

\subsubsection{Dependencies on the Underlying Absolute Magnitude Relation}

\begin{figure}
\epsscale{1.0}
\plottwo{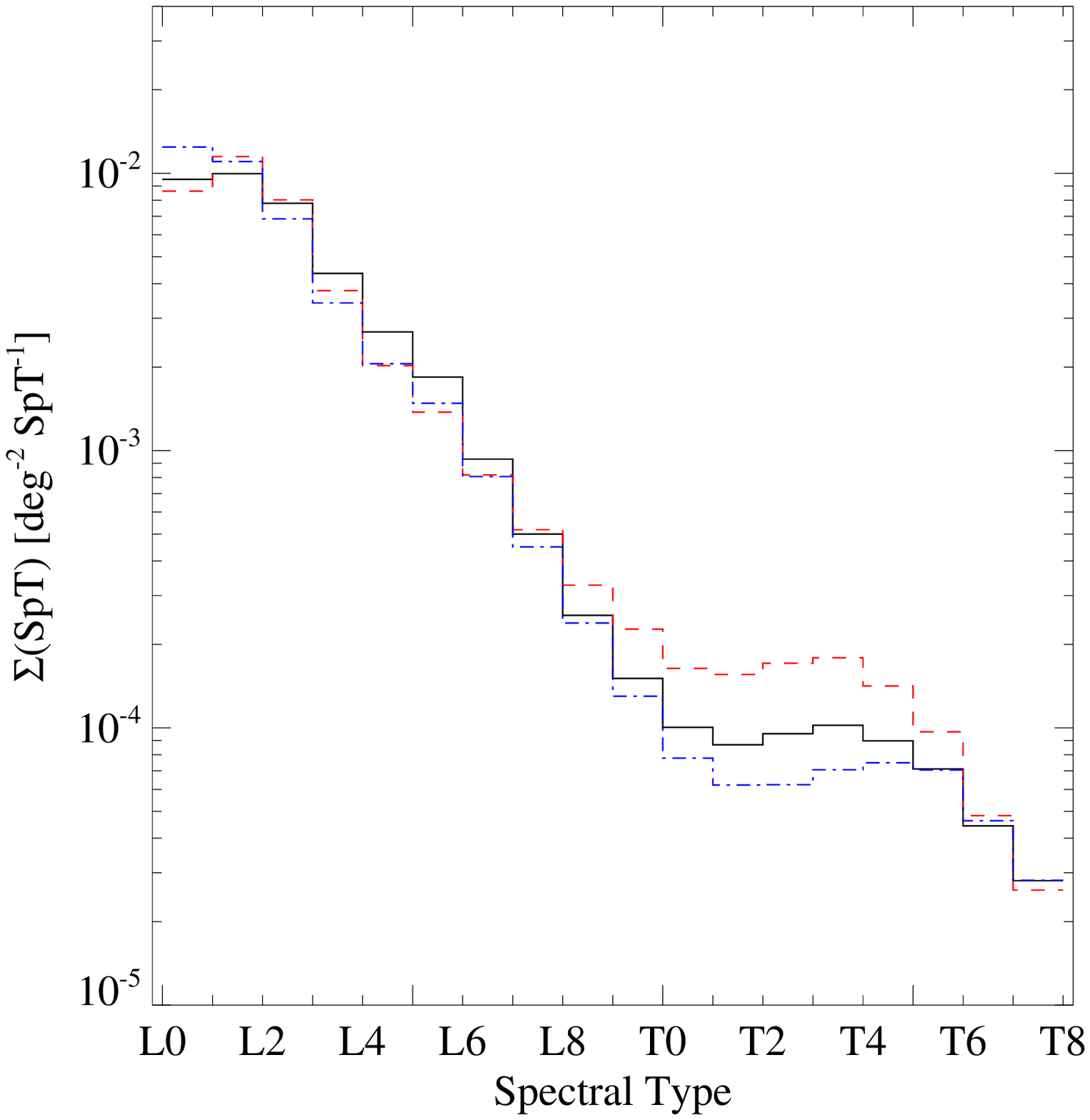}{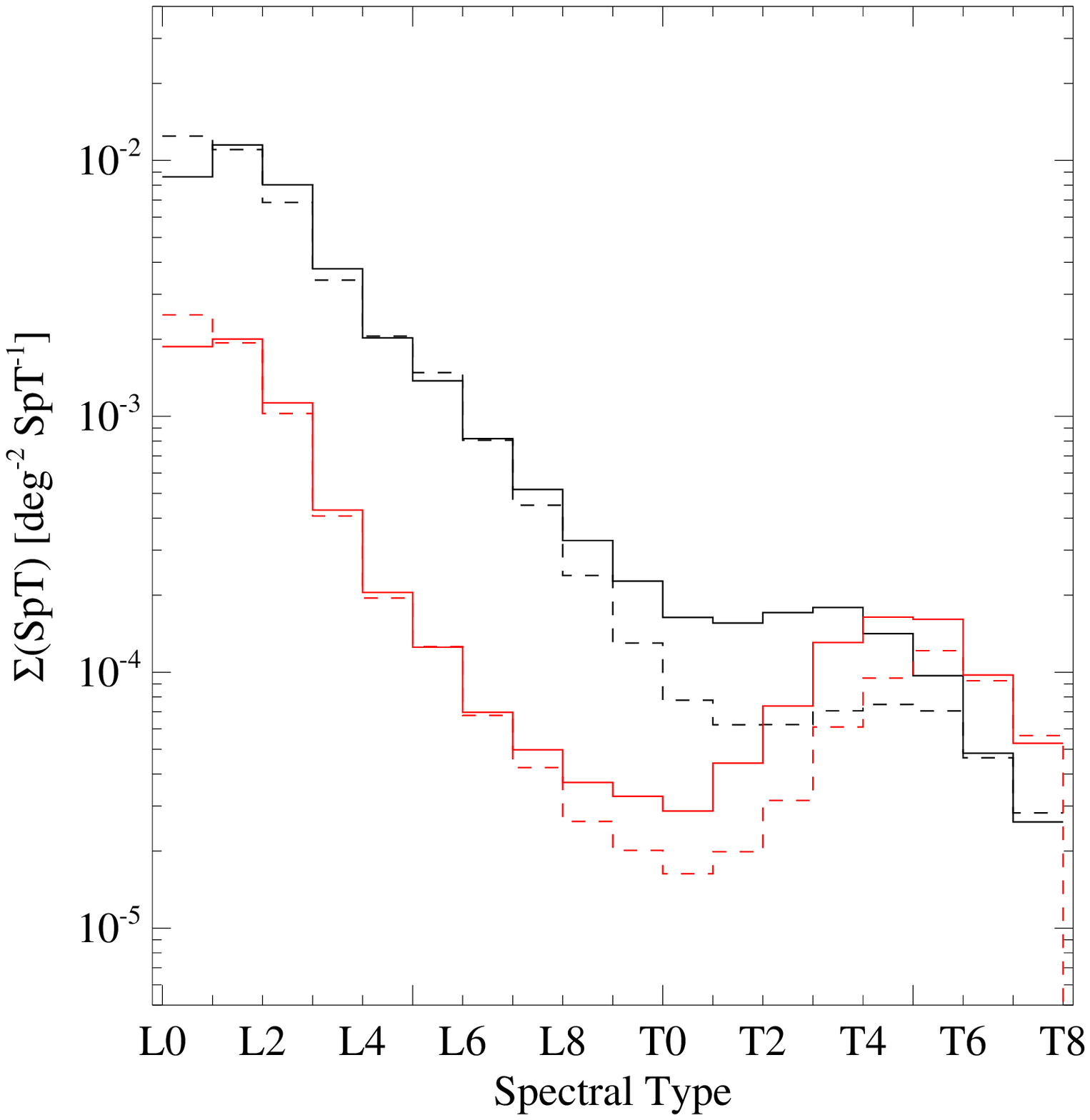}
\caption{Surface density distributions as a function of 
systemic spectral type for different
absolute magnitude/spectral type relations.  (Left panel)
$M_K$/spectral type relations from this study 
(black line; Table~\ref{tab_fits}) and \citet{liu06}, 
excluding known (red dashed line) and known and 
possible binaries (blue dot-dashed line).
(Right panel) Comparison of $M_J$ (red lines) and
$M_K$ (black lines) spectral type relations from \citet{liu06}
excluding known (solid line) and known and 
possible binaries (dashed lines).
All distributions are based on a magnitude-limited sample with 
a limiting magnitude of 16 in their respective
bands.
\label{fig_sptvsabsmag}}
\end{figure}

The spectral type distribution of a magnitude-limited survey
is naturally influenced by the underlying absolute-magnitude 
relation.  As illustrated in Figure~\ref{fig_sptvsabsmag}, 
the most significant differences found in these simulations are 
largely constrained to L/T transition objects, where the 
three absolute magnitude/spectral type relations considered here vary appreciably.  
Most pertinent are the differences between simulations based on 
the two $M_K$/spectral type relations of \citet{liu06}, 
which differ by up to a factor of 2-5 over the L8--T5 range.
This is consistent with the $\sim$1~mag difference
between these relations, 
implying a factor of $\sim$4 increase in volume sampled.  
The large differences in the surface densities imply
that a complete magnitude-limited sample of L/T transition objects
can provide a robust constraint
on the underlying absolute magnitude relation.
Similar results are also found between different $M_J$/spectral
type relations.  
Figure~\ref{fig_sptvsabsmag} also illustrates the differences
between $J$-band limited and $K$-band limited samples.
Surface density distributions in the former case 
show considerably more structure, with more
pronounced 
local minima and maxima at types T0 and T5, respectively.
This reflects the color shift between L dwarfs and T dwarfs
(from red to blue $J-K$ colors), the sharper turnover in 
$M_J$/spectral type relations across the L/T transition
(cf.~Figure~4 in \citet{liu06})
and the increase in number densities beyond T0 (Figure~\ref{fig_sptvsmf}).

\subsubsection{Weakly Correlated Parameters}

Several factors considered in the simulations
contributed minimally to the overall shape of the number density and
surface density distributions.  
These include the different minimum ages for the
field population examined, spanning 0.01 to 1~Gyr, which only served
to decrease the number of early-type L dwarfs by $<$15\%.  This is consistent
with the results of \citet{memf}, which demonstrated that only
extreme modulations of the birthrate change the luminosity function
of brown dwarfs appreciably.  The two evolutionary models examined
also produced similar spectral type distributions, with small 
deviations amongst early-type L dwarfs arising from different
hydrogen-burning minimum masses derived by these models
(0.075~M$_{\sun}$ for Burrows et al.\ 1997 
versus 0.072~M$_{\sun}$ for Baraffe et al.\ 2003).  
Constant and exponential mass ratio distributions produced
nearly identical spectral type distributions for ${\epsilon}_b$ = 0.1,
although 30\% differences in the surface densities of T0--T4 systems
are seen for ${\epsilon}_b$ = 0.5.  
Just as an increase in the underlying binary fraction increases the
total effective volume sampled for a given subclass, the exponential
mass ratio distribution, which favors equal-mass/equal-brightness binaries,
samples a larger effective volume than a flat mass ratio distribution.
However, this effect is considerably smaller than the increase in 
number and surface densities (particularly amongst L/T transition objects)
incurred by an increase in the underlying binary fraction.
While all of these differences between the simulations 
are statistically significant, 
they are considerably smaller
than current empirical uncertainties for L and T dwarf
number and surface density measurements
\citep{me01,cru03}, and can be considered negligible for the
purposes of this study.

\subsection{Binary Fraction Distributions}

\subsubsection{Reproducing the Binary Excess}

\begin{figure}
\epsscale{1.0}
\plottwo{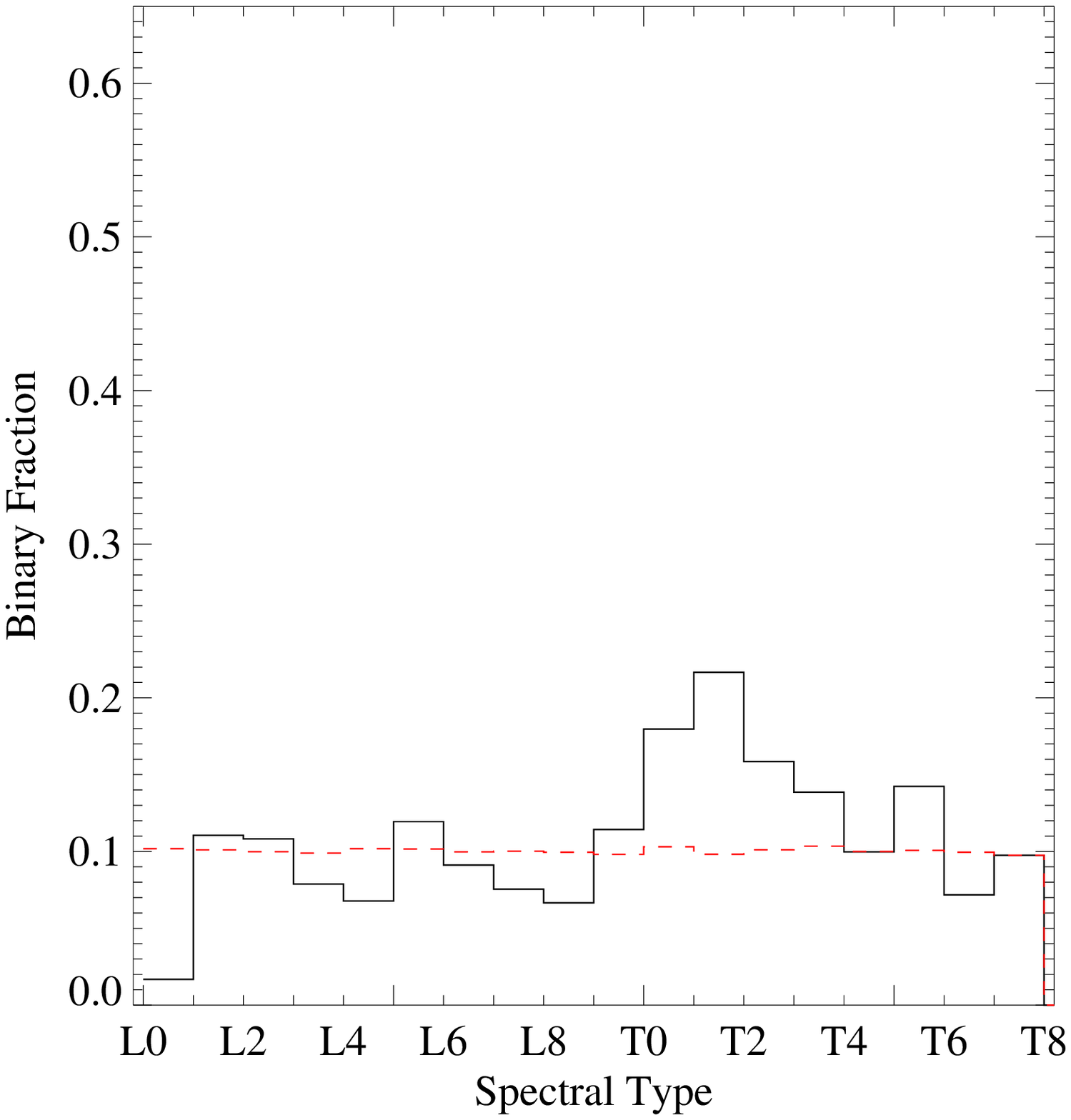}{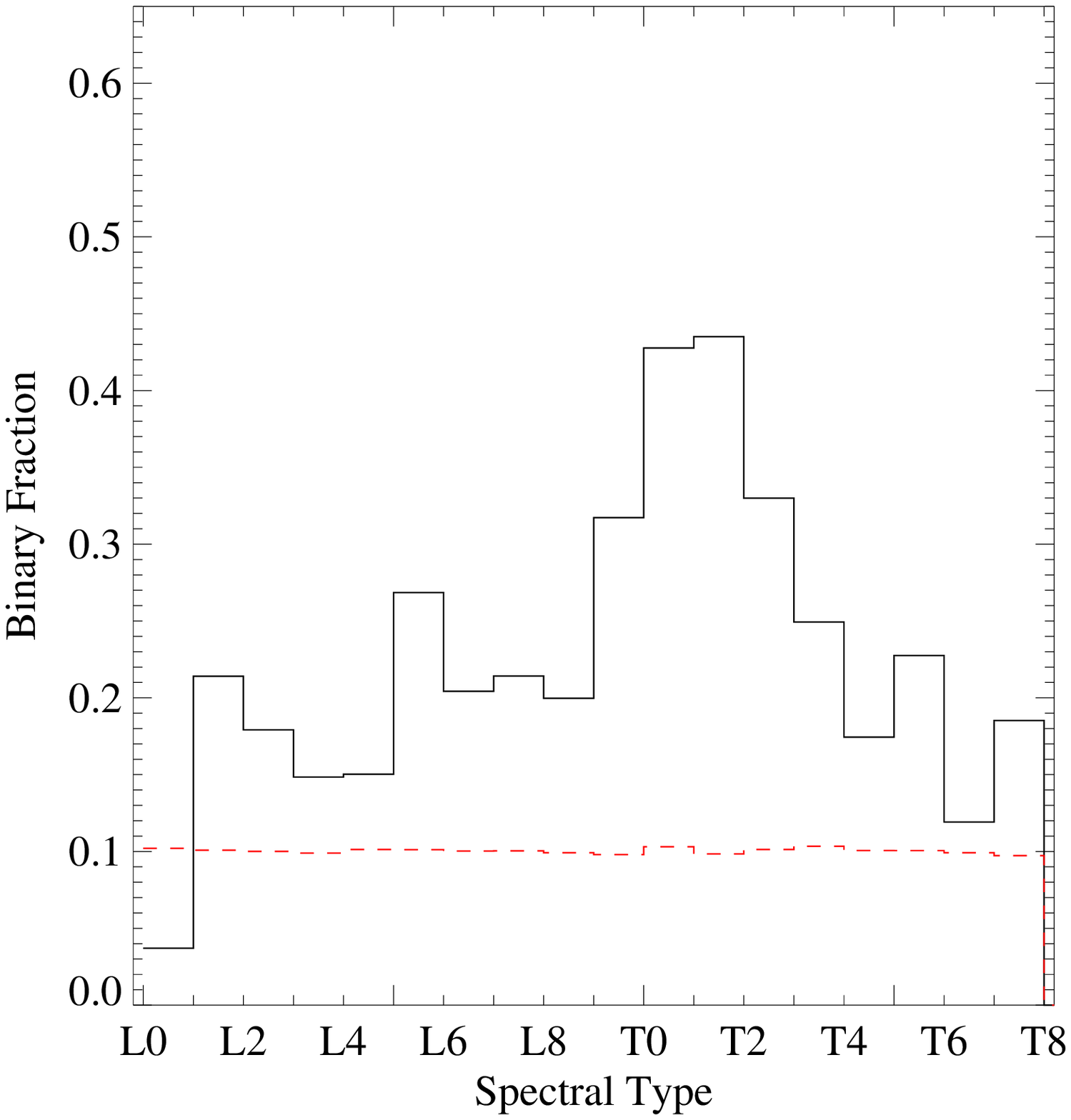}
\caption{Binary fraction distributions
for volume-limited (left) and magnitude-limited samples (right).
In both panels, the binary fraction as a function of systemic spectral
type is indicated in black, while the input binary fraction 
distribution is indicated in red (${\epsilon}_b$ = 0.1).
\label{fig_bfvscomp}}
\end{figure}

Binary fractions as a function of spectral type were computed by 
ratioing the number of binaries within a given systemic
spectral type range to all simulated sources within that same
range. Results from representative simulations for both volume-limited
and magnitude-limited cases are given in 
Tables~\ref{tab_bfdist} and~\ref{tab_bfdistmag}, respectively.
Figure~\ref{fig_bfvscomp} displays the binary fraction
distribution as a function of systemic spectral type for the baseline
simulations.
For both volume-limited and magnitude-limited samples, 
there is a considerable degree of structure in these
distributions, deviating significantly from
the underlying flat form.  
The largest jump is seen amongst T0--T2 dwarfs,
over which binary fractions increase
by a factor of two in the volume-limited case.  
The remaining low-level structure, of order 30--50\%,
likely arises from composite spectral
classification errors\footnote{Simulations with alternate spectral
templates were examined to ascertain whether this
significantly influenced the binary fraction distribution
results.  To the limits of our spectral sample, 
only minimal changes in the 
distribution were found (of order 10\%).
Note that as dips are generally followed or preceded by peaks, 
it is likely that such systematic effects are 
averaged out in the coarser spectral binning used for comparisons
to the empirical data.}
or, in the case of L0--L1 dwarfs, endpoint effects.
However, these are significantly smaller than the
strong peak in the binary fraction
distribution at types T0--T2.

In the magnitude-limited case, all binary fractions are consistently
higher, a well-understood bias incurred by the larger volume sampled by overluminous, unresolved pairs.\footnote{Following
the formalism of \citet{mehst}, a mass ratio
distribution of the form $P(q) \propto q^4$ and intrinsic
binary fraction of 0.1 implies an apparent binary fraction of 
0.19, a factor of roughly two increase.}  For the most part, 
this bias merely serves to
enhance the structure seen in the volume-limited distribution and,
most importantly, the fraction of binaries amongst T0--T2 dwarfs,
which rises to 0.43.  
This result was anticipated by the increase in
the surface density distribution between systems and singles 
as noted in $\S$~4.1.2 (Figure~\ref{fig_sptvscomp}).

That the fraction of binaries amongst T0--T2 dwarfs is higher 
in a volume-limited sample suggests that the {\em intrinsic}
binary fraction of early-type T dwarfs is higher.
However, it is important to make a distinction between the 
binary fraction distribution
of systems and that of primaries.
The latter exhibit a flat distribution
in both the volume-limited and magnitude-limited cases, 
consistent with the assumed fixed underlying binary fraction.
A more appropriate conclusion is that 
early-type T dwarf {\em systems} are
more frequently binary, when such systems are unresolved.

\subsubsection{Comparison to Empirical Data and Constraints on the Underlying Binary Fraction}

\begin{figure}
\epsscale{1.0}
\plotone{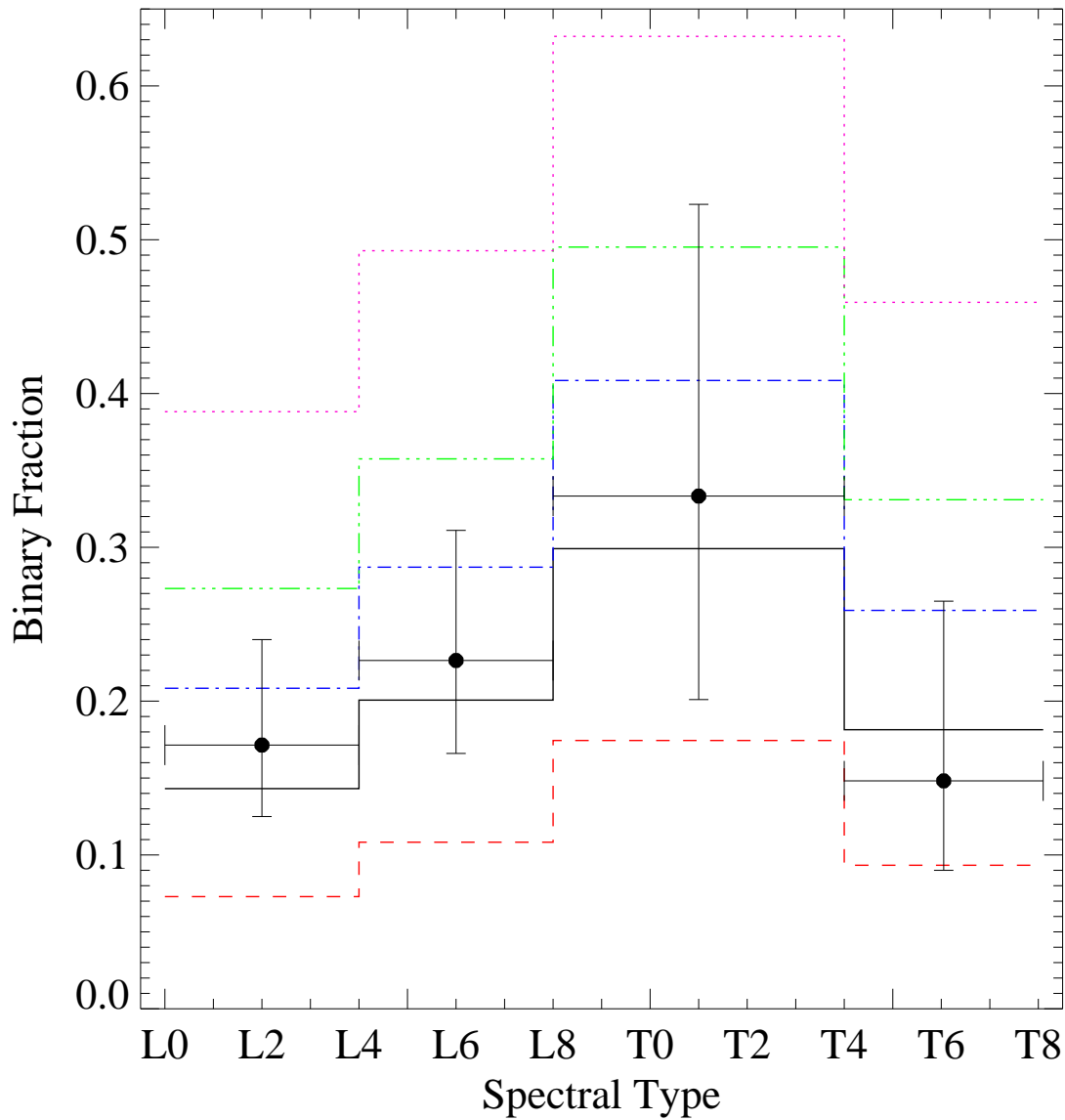}
\caption{Binary fraction distributions for a magnitude-limited sample with
${\epsilon}_b$ = 0.05 (red dashed line), 0.1 (black line), 0.15
(blue dot-dashed line), 0.2 (green triple-dot
dashed line) and 0.3 (magenta dotted line), resampled into
spectral type bins of L0--L4, L4--L8, L8--T4, and T4--T8.
Empirical measurements are indicated by solid circles
with error bars, where the uncertainties 
correspond to the 90\% confidence intervals for a binomial
distribution \citep{mehst}.
\label{fig_bfvsdata}}
\end{figure}

Figure~\ref{fig_bfvsdata} compares the magnitude-limited
binary fraction distribution as a function of systemic 
spectral type to
the binned empirical data of Table~\ref{tab_bindata2}.  
The agreement between these for the baseline parameters is quite 
remarkable.  The steady rise in the binary fraction
between early-type L dwarfs and L/T transition
objects, the peak of the distribution in the latter case, and the
subsequent drop 
in the binary fraction for the latest-type T dwarfs are all faithfully
reproduced in both form and magnitude.

Figure~\ref{fig_bfvsdata} also illustrates how the binary
fraction distribution varies with the assumed underlying
binary fraction, ${\epsilon}_b$.  It is seen that while increasing
${\epsilon}_b$ increases the scale of the distribution, it has
little effect on its shape; in particular, the relative number of
binaries between L/T transition objects and 
early-type L or later-type T dwarfs. 

Comparison between these distributions and the empirical data 
enable a constraint on ${\epsilon}_b$.  
By linearly interpolating the distributions 
as a function of ${\epsilon}_b$
and minimizing the uncertainty-weighted deviations between the data
and simulation results, a best fit ${\epsilon}_b^{res}$ = 0.11 was found, with 
ranges of 0.09--0.15 (0.08--0.17) acceptable at the 90\% (95\%) confidence
levels.  These values are fully consistent with earlier estimates
of the volume-limited, resolved binary fraction of L and T dwarfs, 
also ranging over 0.09--0.15 \citep{bou03,mehst,mehst2,rei06}.  
It is important to remember that ${\epsilon}_b^{res}$ is the
underlying {\em resolved} binary fraction, having been constrained by
observations of resolved systems.  The difference between this and the 
{\em intrinsic} binary fraction of brown dwarfs 
is discussed further in $\S$~5.2.

\subsubsection{Dependencies on the Mass Ratio Distribution}

\begin{figure}
\epsscale{1.0}
\plottwo{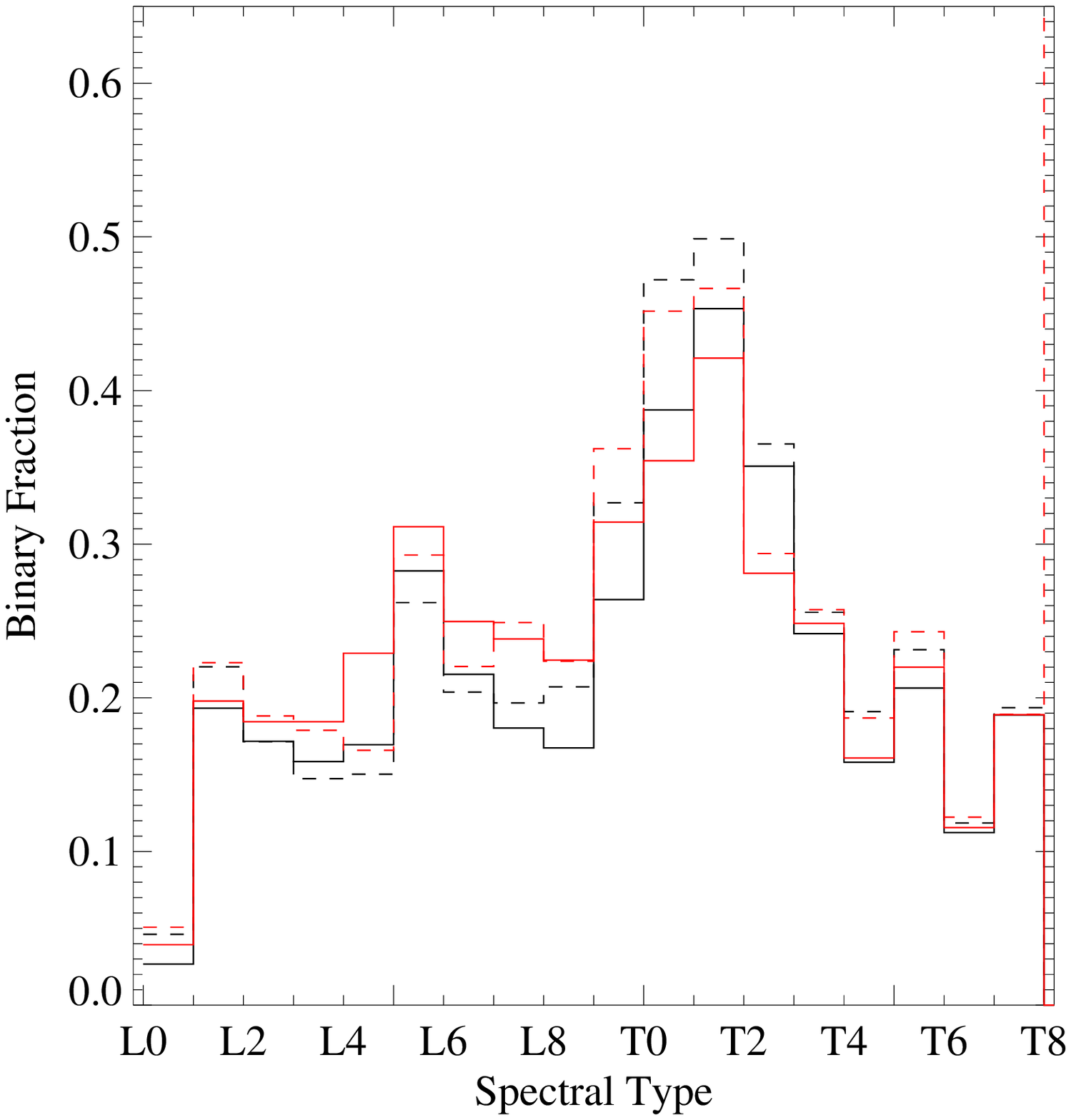}{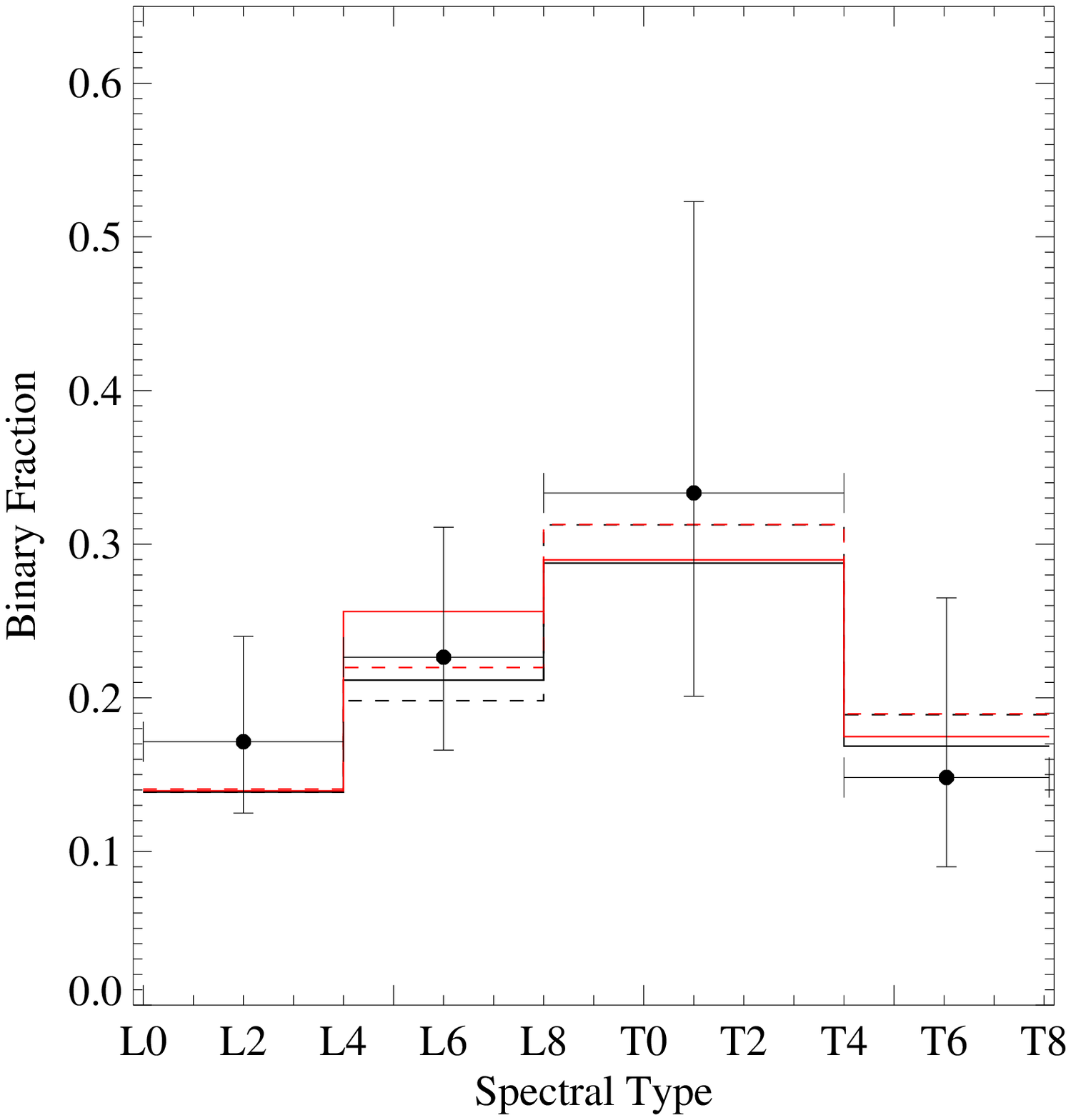}
\caption{Binary fraction distributions for magnitude-limited samples
assuming different forms of the \citet{liu06}
$M_K$ (black lines) and $M_J$
(red lines) spectral type relations, excluding known (solid lines)
or known and possible binaries (dashed lines).
The left panel shows the distributions sampled by integer spectral
type; the right panel shows the same distributions sampled by the
spectral bins listed in Table~\ref{tab_bindata2}.
Empirical measurements are indicated in the right panel
as in Figure~\ref{fig_bfvsdata}.
\label{fig_bfvsabsmag}}
\end{figure}

\begin{figure}
\epsscale{1.0}
\plottwo{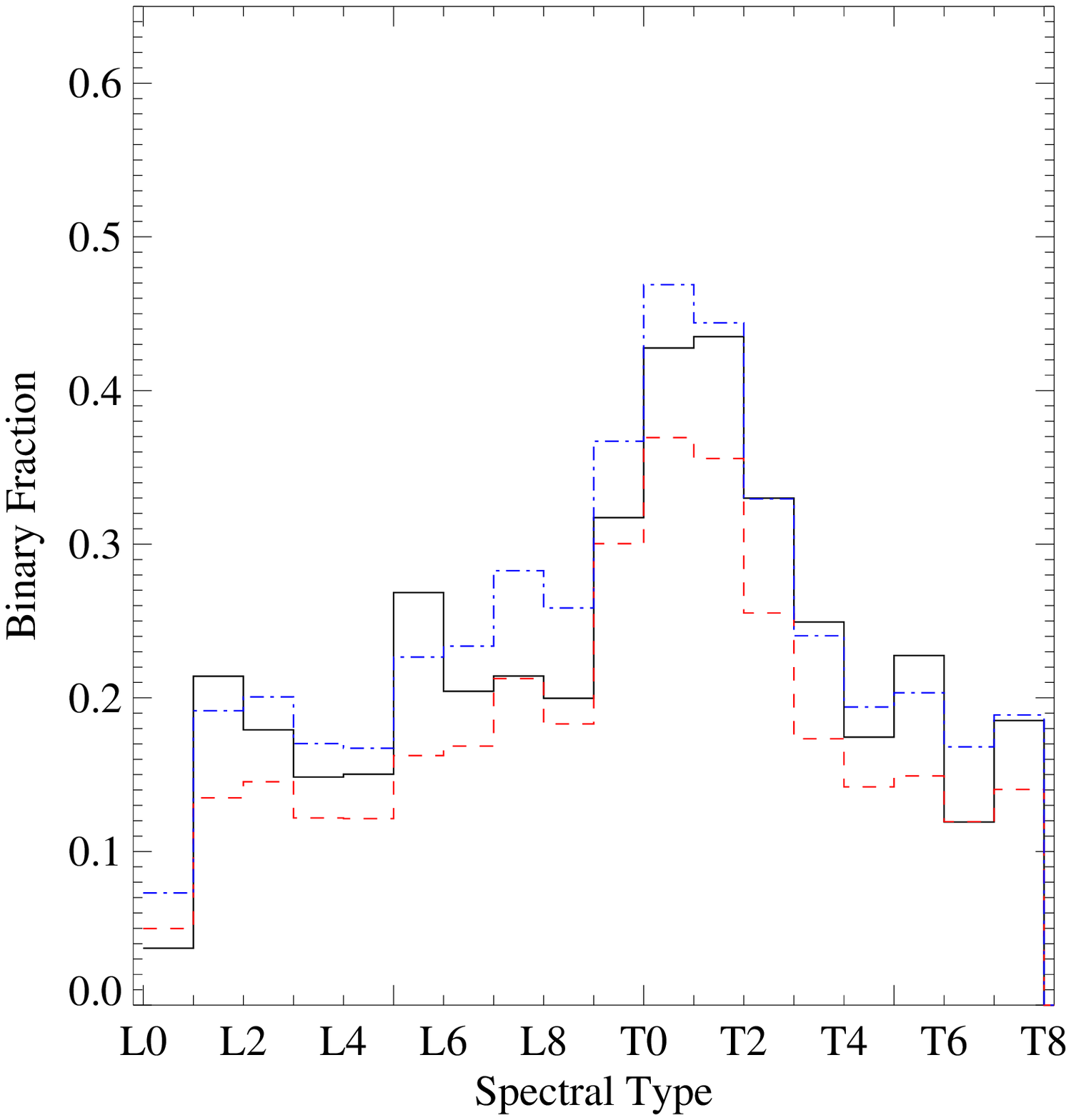}{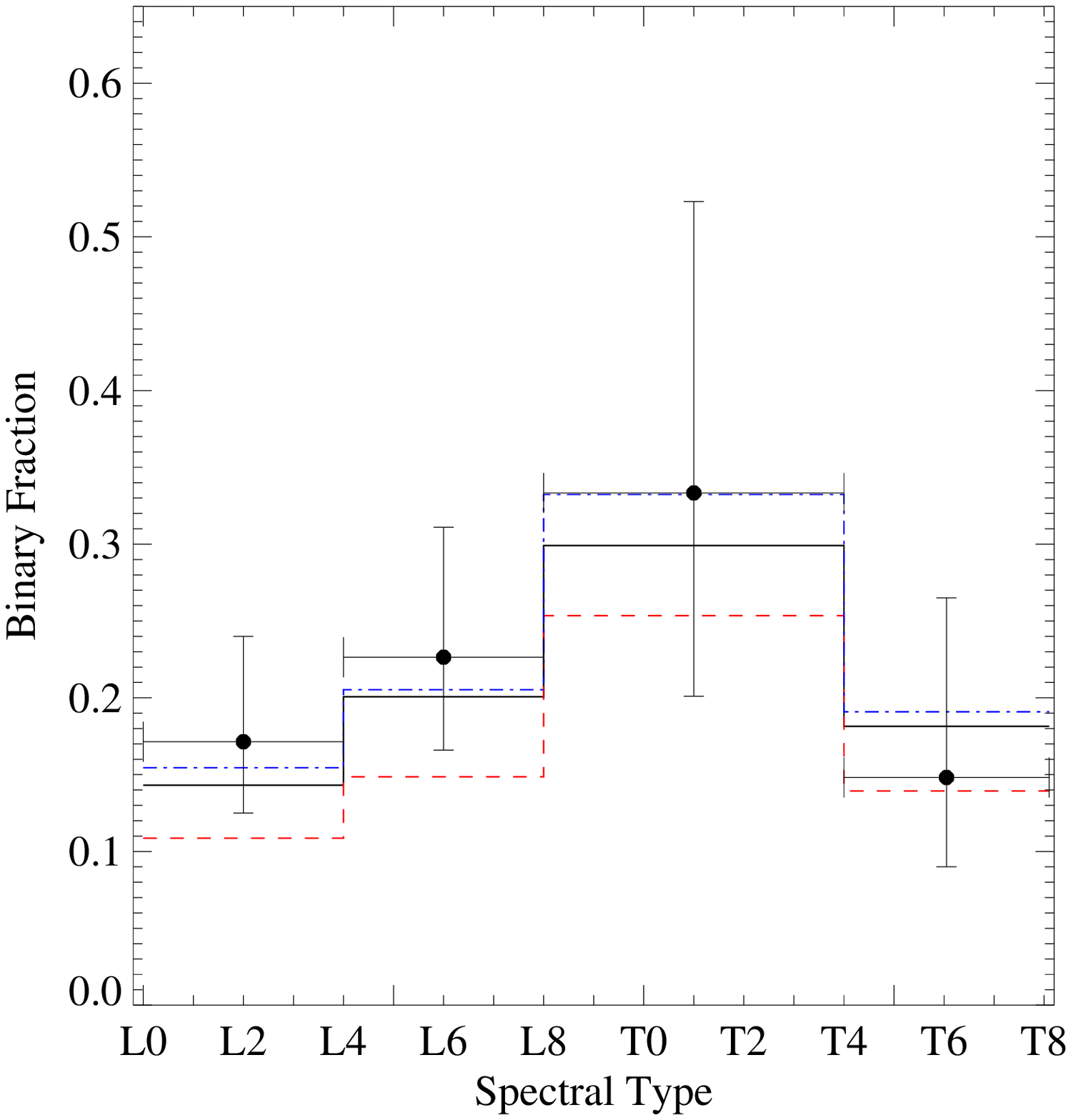}
\caption{Binary fraction distributions for magnitude-limited samples
assuming different forms of the underlying mass ratio distribution.
Black lines plot distributions assuming 
${\epsilon}_b$ = 0.1 and $P(q) \propto q^4$,
red dashed lines plot distributions assuming
${\epsilon}_b$ = 0.1 and $P(q) \propto$ constant,
and blue dot-dashed lines plot distributions assuming
${\epsilon}_b$ = 0.14 and $P(q) \propto 1$.
The left panel shows the binary fraction distributions 
sampled by integer spectral
type; the right panel shows the same distributions sampled by the
spectral bins listed in Table~\ref{tab_bindata2}.
Empirical measurements are indicated in the right panel
as in Figure~\ref{fig_bfvsdata}.
\label{fig_bfvsqdist}}
\end{figure}

A somewhat larger modulation of the apparent binary 
fraction distribution is seen 
when different mass ratio distributions are considered,
as illustrated in Figure~\ref{fig_bfvsqdist}.  
A flat mass ratio distribution results in a 20-30\% decline in the binary
fraction distribution across the full spectral type range,
in both the binned and unbinned cases.  Again, this decline
is consistent with the reduction in effective volume sampled
by the reduced fraction of equal-mass/equal-brightness systems
for a magnitude-limited survey.  
Figure~\ref{fig_bfvsabsmag} illustrates that a simulation
using a flat mass ratio distribution and ${\epsilon}_b$ = 0.14 
produces a binary fraction distribution that
differs by less than 10\% from the baseline simulation.
Indeed, the empirical data constrain a best fit
${\epsilon}_b^{res}$ = 0.14 and acceptable
ranges of 0.12--0.19 (0.11--0.21) at the 90\% (95\%)
confidence limits in the case of
a flat mass ratio distribution, somewhat higher than
that derived using an exponential distribution.\footnote{Following the
formalism of \citet{mehst}, a flat mass ratio distribution
and an inherent binary fraction of 0.1 implies an apparent
binary fraction of 0.14, a decrease of 26\% as compared to the
exponential mass ratio distribution.}
This degeneracy between the mass ratio
distribution and the underlying 
binary fraction implies that the former
cannot be robustly constrained by the binary fraction
distribution without an independent determination
of ${\epsilon}_b$.

\subsubsection{Dependencies on the Absolute Magnitude/Spectral Type Relation}

Figure~\ref{fig_bfvsabsmag} illustrates how the
binary fraction distribution varies for different underlying absolute
magnitude distributions.
Despite the substantial effect
that the absolute magnitude relation has on the surface
density distribution of L8--T5 dwarfs (Figure~\ref{fig_sptvsabsmag}),
the impact on the binary fraction distribution is considerably smaller.
Between the L06 and L06p $M_K$/spectral type relations, 
binary fraction distributions differ by
less than 20\% for individual subtypes,
and by less than 10\% when the spectral types are binned according
to the empirical data.
There is a comparably small difference between binary fraction
distributions generated by the $M_J$/spectral type relations of 
\citet{liu06}, and also between $J$- and $K$-band magnitude-limited samples.
These comparisons indicate that the binary fraction distribution,
and more importantly the binary excess at the L/T transition,
are relatively insensitive to the details of how a magnitude-limited
sample is constructed.  This justifies the incorporation of several
high-resolution imaging samples into a single
magnitude-limited probe of the binary fraction distribution.

A more important conclusion to draw from these comparisons, however,
is that the observed binary excess does not appear to be
an artifact of magnitude-limited imaging samples, since 
the parameters that most influence the surface densities of L
and T dwarfs (the absolute magnitude scale and limiting filter band)
have little bearing on the binary fraction distribution.  
Indeed, as already pointed out in $\S$~4.2.1, a magnitude-limited
sample simply amplifies the excess present in the underlying
population.

\subsubsection{Weakly Correlated Parameters}

As with the number and surface density distributions, several of the
parameters explored in these simulations had minimal influence
on the binary fraction distribution.
%In addition to the small influence of the absolute magnitude relation,
%and the slight degeneracy between the mass ratio distribution and ${\epsilon}_b$,
Neither the form of the mass function, choice of evolutionary model
or minimum age of the population had more than a 5\% effect
on the structure or magnitude of the distribution. 
Out of all of the parameters varied, only ${\epsilon}_b$
significantly influences the binary fraction distribution
and only by scaling the entire distribution up or down.
The structure of the distribution --- i.e., the excess 
of binaries amongst L/T
transition objects --- must arise from some other aspect of the
simulation.  As discussed below, it appears that the 
luminosity scale is exclusively responsible for this excess.

\section{Discussion}

\subsection{On the Origin of the L/T Transition Binary Excess}

The simulations presented here reproduce remarkably
the relatively high binary fraction of L/T transition
brown dwarfs.
Yet how does this binary excess arise?  Both \citet{mehst2}
and \citet{liu06} proposed that the similarity of the
composite spectra of L + T dwarf binaries to early-type
T dwarf spectral morphologies, and the rapid evolution of
single brown dwarfs through the L/T transition, may be the
underlying causes.  A detailed examination of
the simulations presented here support these hypotheses.

\begin{figure}
\epsscale{1.0}
\plotone{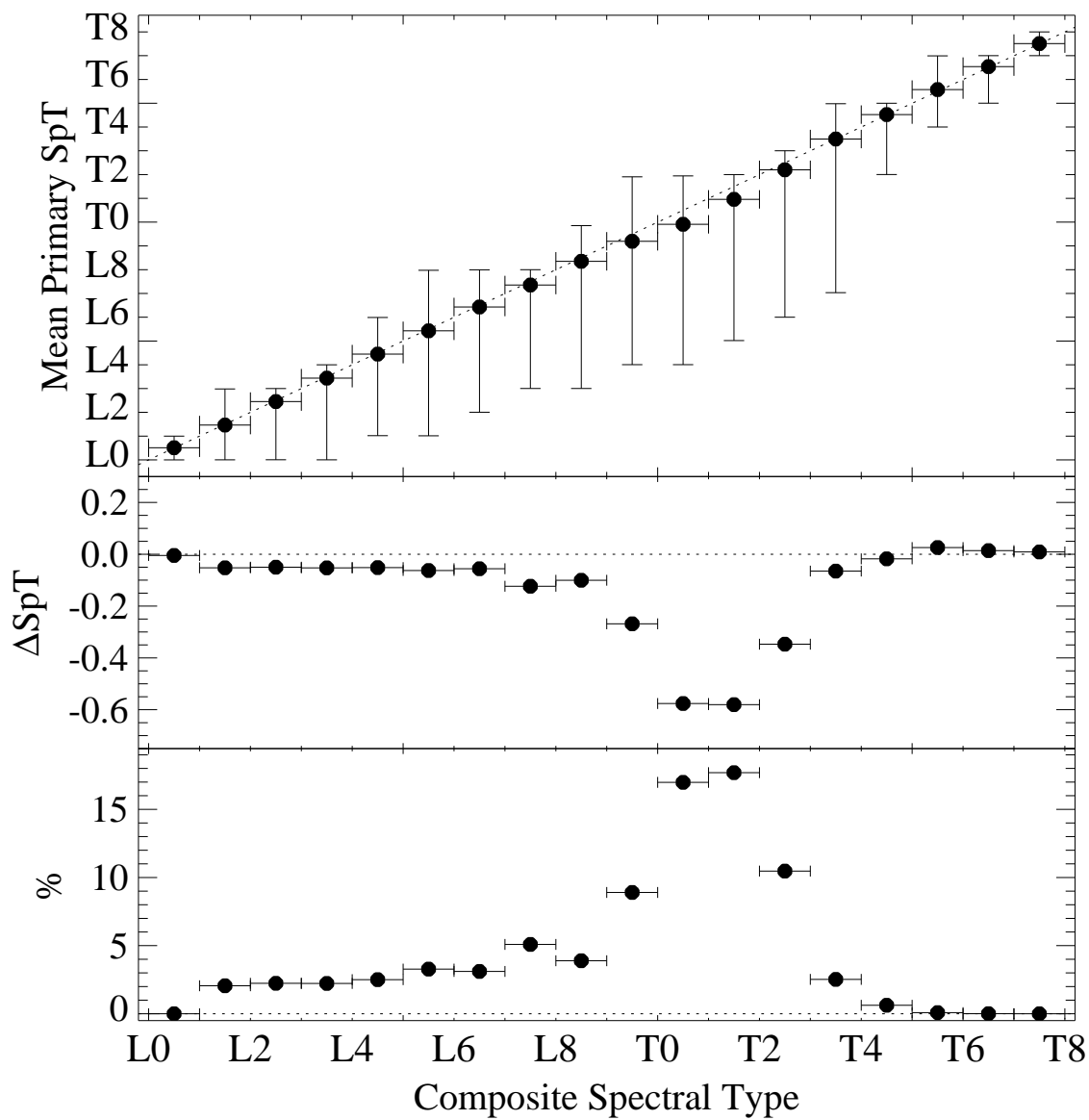}
\caption{Composition of binaries as a function of composite spectral type.
(Top panel) Mean primary spectral type, where the vertical 
error bars indicate the minimum and maximum primary spectral type
of the binaries for a given composite spectral type.
(Middle panel) Average difference between composite and primary spectral type.
(Bottom panel) Percentage of binaries with primaries classified
a full spectral class or earlier than the composite type. 
\label{fig_primspt}}
\end{figure}

First, the composition of L/T transition binaries 
in the simulations is quite distinct as compared to other spectral
subclasses,
as illustrated in Figure~\ref{fig_primspt}.  
Most binaries have composite types
similar to their primaries, reflecting both the preference for 
equal-mass/equal-luminosity systems in the exponential
mass ratio distribution and
the dominance of the primary in the combined light
spectrum of systems with low-mass companions.
Yet binaries classified as L9--T3 dwarfs consistently 
have earlier-type primaries, up to $\sim$0.6 subclasses on average, 
while 15-20\% of T0--T2 systems have primaries at least
one subclass earlier. 
This is consistent with studies of known L/T transition binaries,
70\% of which have been shown or inferred
to be L dwarf plus T dwarf pairs \citep{mehst2}.
Such pairs readily mimic the spectral appearance of
an early-type T dwarf. 
The flattening of the absolute magnitude scale across
the L/T transition allows the T dwarf secondaries of these systems
to contribute significantly to the overall spectral flux at near
infrared wavelengths. Thus, a wide variety of component combinations
result in a composite spectral energy distribution similar to
an early-type T dwarf (cf.~Figure~\ref{fig_hybrid}).

Second, a rapid evolution of brown dwarfs
across the L/T transition can be deduced from the flattening of
the luminosity scale over this range.
Based on the evolutionary models
of \citet{bar03} and the $M_{bol}$/spectral type
relation shown in Figure~\ref{fig_mbol}, a 0.05~M$_{\sun}$ (0.03~M$_{\sun}$) brown dwarf 
dims from type L8 to type T3 in $\sim$600~Myr ($\sim$100~Myr), a period
spanning 30\% (20\%) of its lifetime at that stage.  In comparison, a brown
dwarf of similar mass spends 2-3 times longer cooling from L0 to L8, and 
a full 7~Gyr (3~Gyr) cooling thereafter to the end of
the T spectral class.  
Since single brown dwarfs spend relatively little time 
at the L/T transition, they are much rarer than earlier-type and later-type
brown dwarfs in a given volume (Figure~\ref{fig_sptvsmf}).
Yet, combinations of these other spectral types, even for small
binary fractions, can be comparable in number to single L/T transition
objects, resulting in the perceived binary excess.

These two explanations for the high rate of binaries amongst
L/T transition objects are in fact related to a single
underlying cause: the flattening of the luminosity scale.  
The small decline in $M_{bol}$ across the L/T transition
directly translates
into small changes in the absolute magnitude scale over this range,
and is the root cause of the rarity of single L/T transition objects.  
Consider then the reverse argument:
{\em The observed excess of binaries
at the L/T transition is further evidence of
a flattening in the luminosity scale.}  Indeed, the agreement
between the shape of the simulated binary fraction distribution 
and empirical data, which is largely independent of all other parameters
besides the $M_{bol}$/spectral type relation, suggests that 
the luminosity 
scale of brown dwarfs across the L/T transition is indeed quite
flat, and that brown dwarfs must traverse this transition over a relatively
short period. 
This lends support to suggestions that the complete removal of
photospheric condensate dust between late-type L and mid-type T, which appears
to be largely responsible for the changes in spectral morphology across
this transition, is 
relatively rapid and may require dynamic,
nonequilibrium atmospheric processes.

\subsection{Absolute Magnitudes of Field L/T Dwarfs and the Intrinsic Brown Dwarf Binary Fraction}

In $\S$~4.2.2, a
constraint on the underlying resolved binary fraction, ${\epsilon}_b^{res}$,
was made by comparing simulated binary fraction distributions
to high-resolution
imaging results.  However, imaging cannot identify
binaries that have very small separations or
systems observed when the components are aligned along the line of sight
(e.g., Kelu~1AB; \citet{mar99,liu05,gel06}).  
The existence of very tight brown dwarf binaries in
the field is likely, as the
separation distribution of resolved systems peaks at or near the
resolution limit of current imaging surveys \citep{bou03,giz03,mehst2,meppv}.
\citet{max05}, analyzing the results of the 
first searches for spectroscopic brown dwarf binaries
\citep{bas99,gue03,ken05,joe06}, have suggested that the true
binary fraction of brown dwarfs may be as high as 0.45 when
selection effects are considered.  
Similarly, \citet{pin03} and \citet{cha05} have inferred 
binary fractions of 0.3--0.5 for very low
mass stars and brown
dwarfs in two young
open clusters based on the identification of overluminous
sources in the color-magnitude diagram.

For field brown dwarfs, particularly L/T transition objects, 
the identification of overluminous,
unresolved binaries is made difficult by uncertainties
in the absolute magnitude relation,
as well as intrinsic scatter arising from age, surface gravity and radius
differences.  Nevertheless, \citet{liu06} hypothesized that
a substantial fraction of L7--T5 dwarfs
with parallax distance measurements 
--- 10 of 15 sources (0.66$^{+0.12}_{-0.17}$\%) --- 
are either resolved or unresolved binaries, due to their outlying
position on the {\teff}/spectral type relation of \citet{gol04}.  
Taking this suggestion at face value, and assuming that
the current parallax sample is essentially magnitude-limited (i.e.,
ignoring additional selection effects),
the fraction of known and possible binaries over this spectral type range  
implies ${\epsilon}_b$ = 0.38, with an acceptable
range of 0.24--0.53 (0.21--0.59) at the 90\% (95\%) confidence level,
assuming\footnote{For a flat
mass ratio distribution, ${\epsilon}_b$ = 0.42, with 
0.27--0.57 (0.24--0.62) acceptable at the 90\% (95\%) confidence level.}
 $P(q) \propto q^4$.  This is substantially higher
than the fraction inferred from high-resolution imaging
surveys, and suggests that up to twice as many 
brown dwarf binaries remain unresolved in these surveys.
Indeed, since counting overluminous sources includes
only those binaries with similar-mass companions,
the intrinsic binary fraction may be higher still.  
This result is seemingly consistent with the high brown dwarf
binary fraction proposed by \citet{max05}.  

Care must be taken in interpreting this value since the number of
L7--T5 dwarfs with parallax measurements is small,
uncertainties in several of the parallax measurements large,
the underlying absolute magnitude distribution
poorly constrained
and hence the identification of overluminous sources 
highly uncertain.
%\footnote{\citet{liu06} state that their
%assumption that all T2--T4.5 dwarfs with parallax measurements
%are binary is ``an extreme example''.}
However, if the absolute magnitude scale is relatively flat
across the L/T transition, and the apparent fraction of
binaries high, a measure of the fraction of overluminous
sources in this spectral type regime may provide a more robust
constraint on the true binary fraction of all brown dwarfs than
high-resolution imaging surveys.

\section{Summary}

The primary results of this study are as follows:

\begin{itemize}
\item The binary fraction excess observed amongst L/T transition objects has
been successfully reproduced by mass function simulations incorporating
the construction and classification of unresolved binary systems.
It is found
that this excess arises largely from a 
flattening in the $M_{bol}$/spectral type relation,
which significantly reduces the number of single L/T transition objects
in a given volume while also causing L dwarf + T dwarf pairs to
mimic the spectral properties of early-type T dwarfs.
While the binary excess is
particularly pronounced in magnitude-limited samples (hence its
detection in current high resolution imaging studies), it is
not caused by selection biases, but is instead
intrinsic to any sample of unresolved brown dwarf
systems.
\item The shape of the binary fraction distribution depends weakly
on the underlying absolute magnitude/spectral type relation,
mass function and minimum age of the field population
(up to 1~Gyr).  The underlying binary fraction, ${\epsilon}_b$ 
produces the greatest effect, scaling the distribution but
not significantly affecting its shape.  There is a slight degeneracy between
the influence of ${\epsilon}_b$ and the mass ratio distribution on
the scale of the binary fraction distribution for a magnitude-limited
sample.
\item The surface density of L/T transition objects in 
a magnitude-limited sample
is highly sensitive to the underlying absolute magnitude/spectral type
relation.  Number and surface densities also scale
with the underlying mass function, but are largely insensitive to the
minimum age of the field population or the mass ratio distribution
of brown dwarf binaries. 
\item Empirical results constrain the underlying resolved
binary fraction to 11$_{-3}^{+6}$\% (90\% confidence interval), 
consistent with prior estimates. The true binary fraction, 
which includes
sources unresolved in imaging studies, may be as high as 40\%
based on the fraction of apparently
overluminous L7--T5 dwarfs as proposed by \citet{liu06}.  However, this 
result requires a more robust assessment of the absolute
magnitude scale, and understanding a selection effects
in the current parallax sample, and more (and improved) 
parallax measurements of L/T
transition objects.
\end{itemize}

The simulations presented here provide new insight into the relationship
between the fundamental properties of brown dwarfs (luminosities, mass
function,  
intrinsic multiplicity) and the observed density, spectroscopic 
properties and multiplicity of L/T transition dwarfs.  
What they do not reveal is the physical mechanism that drives
this transition.  The dramatic change in spectral properties
over a short evolutionary period
is strong evidence that nonequilibrium
dispersion of photospheric condensates must play
an important, if not predominant, role.  As new theoretical models
address dynamical atmospheric effects that may be involved in 
condensate cloud evolution, continued empirical characterization
of these sources --- including improved multiplicity statistics
through high-resolution imaging, spectroscopic monitoring,
and parallax measurements ---
should be a priority.
Fortunately, the apparently high binary fraction of
L/T transition objects
implies that this remarkable evolutionary phase of brown dwarfs
can be studied quite thoroughly, through resolved
photometric and spectroscopic studies and long-term 
astrometric monitoring to measure component masses.
These enigmatic systems may ultimately provide the most stringent
empirical constraints on the properties of brown dwarfs in general. 

\acknowledgments

The author thanks the Infrared Telescope Facility
telescope operators and instrument specialist John Rayner
for their assistance in the acquisition of the SpeX data
presented here; and Kelle Cruz for providing 
additional SpeX data used in this study.
He also thanks the referee, Patrick Lowrance, for his helpful
review of the original manuscript; and Adam Burrows and
Michael Liu for their comments.
This publication makes
use of data from the Two Micron All Sky Survey, which is a joint
project of the University of Massachusetts and the Infrared
Processing and Analysis Center, and funded by the National
Aeronautics and Space Administration and the National Science
Foundation. 2MASS data were obtained from the NASA/IPAC Infrared
Science Archive, which is operated by the Jet Propulsion
Laboratory, California Institute of Technology, under contract
with the National Aeronautics and Space Administration.
This program has benefitted from the M, L, and T dwarf compendium housed at DwarfArchives.org and maintained by Chris Gelino, Davy Kirkpatrick, and Adam Burgasser;
and the VLM Binary Archive maintained by N. Siegler at 
\url{http://paperclip.as.arizona.edu/$\sim$nsiegler/VLM\_binaries/}.
The authors wish to recognize and acknowledge the 
very significant cultural role and reverence that 
the summit of Mauna Kea has always had within the 
indigenous Hawaiian community.  We are most fortunate 
to have the opportunity to conduct observations from this mountain.

Facilities: \facility{IRTF(SpeX)}

\clearpage

\begin{deluxetable}{lccccc}
\tabletypesize{\footnotesize}
\tablecaption{Empirical Binary Fraction Data \label{tab_bindata2}}
\tablewidth{0pt}
\tablehead{
\colhead{SpT} &
\colhead{\# Total} &
\colhead{\# Binary} &
\colhead{${\epsilon}_b^{obs}$} &
\colhead{90\% CI\tablenotemark{a}} &
\colhead{95\% CI\tablenotemark{a}}  \\
}
\startdata
L0--L0.5 & 15 & 4 & 0.27 & 0.16--0.44 & 0.13--0.48 \\ 
L1--L1.5 & 20 & 2 & 0.10 & 0.05--0.23 & 0.04--0.27 \\ 
L2--L2.5 & 22 & 3 & 0.14 & 0.08--0.27 & 0.06--0.30 \\ 
L3--L3.5 & 13 & 3 & 0.23 & 0.13--0.42 & 0.11--0.47 \\ 
L4--L4.5 & 18 & 2 & 0.11 & 0.06--0.26 & 0.05--0.30 \\ 
L5--L5.5 & 17 & 3 & 0.18 & 0.10--0.33 & 0.08--0.38 \\ 
L6--L6.5 & 10 & 3 & 0.30 & 0.17--0.51 & 0.14--0.56 \\ 
L7--L7.5 & 8 & 4 & 0.50 & 0.30--0.70 & 0.25--0.75 \\ 
L8--L8.5 & 3 & 0 & 0.00 & $<$0.44 & $<$0.53 \\ 
L9--L9.5 & 1 & 0 & 0.00 & $<$0.68 & $<$0.78 \\ 
T0--T0.5 & 2 & 2 & 1.00 & $>$0.47 & $>$0.37 \\ 
T1--T1.5 & 3 & 1 & 0.33 & 0.14--0.68 & 0.10--0.75 \\ 
T2--T2.5 & 1 & 0 & 0.00 & $<$0.68 & $<$0.78 \\ 
T3--T3.5 & 2 & 1 & 0.50 & 0.20--0.80 & 0.14--0.86 \\ 
T4--T4.5 & 4 & 1 & 0.25 & 0.11--0.58 & 0.08--0.66 \\ 
T5--T5.5 & 8 & 1 & 0.13 & 0.06--0.37 & 0.04--0.43 \\ 
T6--T6.5 & 8 & 1 & 0.13 & 0.06--0.37 & 0.04--0.43 \\ 
T7--T7.5 & 6 & 1 & 0.17 & 0.08--0.45 & 0.05--0.52 \\ 
T8--T8.5 & 1 & 0 & 0.00 & $<$0.68 & $<$0.78 \\ 
\cline{1-6}
L0--L3.5 & 70 & 12 & 0.17 & 0.13--0.24 & 0.11--0.26 \\
L4--L7.5 & 53 & 12 & 0.23 & 0.17--0.31 & 0.15--0.34 \\
L8--T3.5 & 12 & 4  & 0.33 & 0.20--0.52 & 0.17--0.57 \\
T4--T8 & 27 & 4 & 0.15 & 0.09--0.27 & 0.07--0.30 \\
\enddata
\tablenotetext{a}{Confidence intervals based on the binomial distribution
\citep{mehst}.}
\tablecomments{Data compiled from high resolution imaging surveys by \citet{mar99,rei01b,rei06,bou03,mehst,mehst2,clo03,giz03,liu05}; and \citet{gel06}.}
\end{deluxetable}

\begin{deluxetable}{ll}
\tabletypesize{\footnotesize}
\tablecaption{Input Parameters for Monte Carlo Simulations.\label{tab_parameters}}
\tablewidth{0pt}
\tablehead{
\colhead{Parameter} &
\colhead{Form and Details}  \\
}
\startdata
Mass Function & {\bf $\Psi({\rm M}) \propto {\rm M}^{-{\alpha}}$}, with $\alpha$ = 0.0, 0.5\tablenotemark{a}, 1.0, 1.5 \\
 & $\Psi({\rm M}) \propto {\rm e}^{-\frac{({\rm \log{M}}-{\rm \log{M_c}})^2}{2{\sigma}^2}}$, with ${\rm M_c} = 0.1 {\rm M}_{\sun}, \sigma = 0.627$\tablenotemark{b} \\
 & \\
Age Distribution & $P(t) \propto$ constant, with $t \in \{T_0,10~{\rm Gyr}\}$, $T_0$ = 0.1\tablenotemark{a}, 0.5, 1.0~Gyr \\
 & \\
Evolutionary Model & \citet{bar03}\tablenotemark{a}; \citet{bur01} \\
 & \\
${\epsilon}_b$ & 0, 0.05, 0.1\tablenotemark{a}, 0.14, 0.15, 0.2, 0.25, 0.3, 0.35, 0.4, 0.45, 0.5, 0.55, 0.6, 0.7 \\
 & \\
Mass Ratio Distribution & $P(q) \propto q^{\Gamma}$, with $\Gamma$ = 4\tablenotemark{a} \\
 & $P(q) \propto$ constant  \\
 & \\
$M_{bol}$/SpT Relation & This paper\tablenotemark{a} (see Table~\ref{tab_fits}) \\
 & \\
$M_K$/SpT Relation & This paper\tablenotemark{a} (see Table~\ref{tab_fits}) \\
 & \citet{liu06}  \\
 & \\
$M_J$/SpT Relation &  \citet{liu06}  \\
\enddata
\tablenotetext{a}{Parameter used in baseline simulation.}
\tablenotetext{b}{Parameters from \citet{cha02}.}
\end{deluxetable}

\begin{deluxetable}{llcccccccccc}
\tabletypesize{\scriptsize}
\tablecaption{Polynomial Fits to Spectral Type Relations \label{tab_fits}}
\tablewidth{0pt}
%\rotate
\tablehead{
 & & \multicolumn{9}{c}{Coefficients} & \\
\cline{3-11}
\colhead{Parameter} &
\colhead{Range} &
\colhead{$c_0$} &
\colhead{$c_1$} &
\colhead{$c_2$} &
\colhead{$c_3$} &
\colhead{$c_4$} &
\colhead{$c_5$} &
\colhead{$c_6$} &
\colhead{$c_7$} &
\colhead{$c_8$} &
\colhead{$\sigma$}  \\
}
\startdata
$M_{bol}$(SpT) & L0--T8 & 1.374e1  &  1.903e-1 &  1.731e-2 &  7.400e-3 & -1.751e-3 & 1.142e-4 &  -2.322e-6 & \nodata & \nodata &  0.22 mag \\
$M_K$(SpT) & L0--T8 & 1.045e1 &    2.322e-1 &   5.129e-2 &  -4.024e-2 &  1.414e-2 & -2.271e-3 &  1.807e-4 & -6.985e-6 & 1.051e-7 & 0.26 mag  \\
SpT({\wat}-J) & L0--T8 & 1.949e1  &    -3.919e1   &    1.312e2   &   -2.156e2  &     1.038e2 & \nodata &  \nodata &  \nodata &  \nodata &  0.8 SpT  \\
SpT({\meth}-J) & T0--T8 & 2.098e1  &    -1.978e1   &    2.527e1   &   -3.221e1  &    9.087e-1 & \nodata &  \nodata &  \nodata &  \nodata &  0.7 SpT  \\
SpT({\wat}-H) & L0--T8 & 2.708e1  &    -8.450e1    &   2.424e2   &   -3.381e2  &     1.491e2 & \nodata &  \nodata &  \nodata &  \nodata &  1.0 SpT   \\
SpT({\meth}-H) & T1--T8 & 2.013e1  &    -2.291e1    &   4.361e1   &   -5.068e1  &     2.084e1 & \nodata &  \nodata &  \nodata &  \nodata &  0.3 SpT   \\
%SpT({\wat}-K & L0--T8 & 4.705e1  &    -2.219e2    &   6.268e2   &   -8.047e2  &     3.536e2 & \nodata &  \nodata &  \nodata &  \nodata &  0.9 SpT   \\
SpT({\meth}-K) & L0--T7 & 1.885e1  &    -2.246e1    &   2.534e1   &   -4.734e0  &    -1.259e1 & \nodata &  \nodata &  \nodata &  \nodata &  1.1 SpT   \\
\enddata
\tablecomments{Polynomials relations are defined as $f(x) = \sum_i{c_ix^i}$.  SpT is the numerical spectral type defined as
SpT(L0) = 0, SpT(L5) = 5, SpT(T0) = 10, etc.}
\end{deluxetable}

\begin{deluxetable}{lllcccl}
\tabletypesize{\scriptsize}
\tablecaption{Spectral Templates \label{tab_standards}}
\tablewidth{0pt}
\tablehead{
 & \multicolumn{2}{c}{Spectral Type} & 
 \multicolumn{2}{c}{2MASS Photometry} \\
\cline{2-3} \cline{4-5}
\colhead{Name} &
\colhead{Opt} &
\colhead{NIR} &
\colhead{$J$} &
\colhead{$J-K_s$} &
\colhead{$\pi$} &
\colhead{Ref} \\
\colhead{} &
\colhead{} &
\colhead{} &
\colhead{(mag)} &
\colhead{(mag)} &
\colhead{(marcs)} &
\colhead{} \\
\colhead{(1)} &
\colhead{(2)} &
\colhead{(3)} &
\colhead{(4)} &
\colhead{(5)} &
\colhead{(6)} &
\colhead{(7)} \\
}
\startdata
2MASS J03454316+2540233  & L0\tablenotemark{a} & L1$\pm$1 & 14.00$\pm$0.03 & 1.33$\pm$0.04 & 37.1$\pm$0.5  & 1,2,3 \\
2MASS J14392836+1929149 & L1\tablenotemark{a} & \nodata &  12.76$\pm$0.02 & 1.21$\pm$0.03 & 69.6$\pm$0.5  & 2,4   \\
SSSPM J0829-1309 & L2 & \nodata & 12.80$\pm$0.03 & 1.51$\pm$0.04 & \nodata  & 5,6 \\
2MASS J15065441+1321060  & L3 & \nodata & 13.37$\pm$0.02 &   1.62$\pm$0.03  & \nodata  & 7 \\
2MASS J11040127+1959217 & L4\tablenotemark{a}  & \nodata &  14.38$\pm$0.03 & 1.43$\pm$0.04 & \nodata & 8 \\
GJ 1001BC\tablenotemark{b} & L5 & L4.5 & 13.11$\pm$0.02 & 1.71$\pm$0.04  & 77$\pm$4 &  3,9,10,11 \\
2MASS J04390101-2353083 & L6\tablenotemark{c} & \nodata & 14.41$\pm$0.03 & 1.59$\pm$0.04 & \nodata & 8 \\
2MASS J09153413+0422045\tablenotemark{b} & L7 & \nodata & 14.55$\pm$0.03 & 1.54$\pm$0.05 & \nodata  & 12  \\
DENIS J025503.3$-$470049 & L8  & \nodata &  13.25$\pm$0.03 & 1.69$\pm$0.04  & \nodata & 13,14 \\
2MASS J03105986+1648155\tablenotemark{b} & L8 & L9 & 16.03$\pm$0.08 & 1.71$\pm$0.11 & \nodata & 10,15 \\
SDSS J042348.57$-$041403.5\tablenotemark{b} & L7.5 & T0\tablenotemark{d} &  14.47$\pm$0.03 & 1.54$\pm$0.04 & 65.9$\pm$1.7   & 16,17 \\
SDSS J015141.69+124429.6 & \nodata &  T1\tablenotemark{d} & 16.57$\pm$0.13 & 1.38$\pm$0.23 & 47$\pm$3  & 15,16 \\
SDSS J125453.90$-$012247.4 & T2 &  T2\tablenotemark{d} &  14.89$\pm$0.04 & 1.05$\pm$0.06 & 73.2$\pm$1.9  & 16,17,18 \\
2MASS J12095613$-$1004008 & \nodata &  T3\tablenotemark{d} &  15.91$\pm$0.07 & 0.85$\pm$0.16 & \nodata & 19 \\
2MASS J22541892+3123498 & \nodata & T4\tablenotemark{d} &  15.26$\pm$0.05 & 0.36$\pm$0.15 & \nodata & 20 \\
2MASS J15031961+2525196 & \nodata & T5\tablenotemark{d} &  13.94$\pm$0.02 & -0.03$\pm$0.06 & \nodata   & 21 \\
SDSS J162414.37+002915.6 & \nodata & T6\tablenotemark{d} &  15.49$\pm$0.05 & $<$-0.02 & 90.9$\pm$1.2  & 22,23 \\
2MASS J07271824+1710012 & T8 &  T7\tablenotemark{d} & 15.60$\pm$0.06 & 0.04$\pm$0.20 & 110$\pm$2  & 17,20 \\
2MASS J04151954$-$0935066 & T8 &  T8\tablenotemark{d} & 15.70$\pm$0.06 & 0.27$\pm$0.21 & 174$\pm$3  & 17,20 \\
\enddata
\tablenotetext{a}{{L} dwarf optical spectral standard from \citet{kir99}.}
\tablenotetext{b}{Known binary \citep{gol04,me0423,rei06,stu06}.}
\tablenotetext{c}{\citet{cru03} classify this source L6.5; however, it is adopted as an L6 spectral template here.}
\tablenotetext{d}{Primary or secondary T dwarf near-infrared spectral standard from \citet{meclass}.}
\tablerefs{(1) \citet{kir97}; (2) \citet{dah02}; (3) \citet{kna04}; (4) \citet{kir99}; (5) \citet{schz02}; 
(6) \citet{lod05}; (7) \citet{giz00}; (8) \citet{cru03}; (9) \citet{gol99}; (10) \citet{kir00}; 
(11) \citet{hen06}; (12) I.~N.\ Reid et al., in preparation; (13) \citet{mar99b}; 
(14) J.\ D.\ Kirkpatrick et al., in preparation; 
(15) \citet{geb02};
(16) \citet{leg00}; (17) \citet{vrb04};
(18) \citet{me03opt}; (19) \citet{mewide3}; (20) \citet{me02a}; (21) \citet{mewide1}; (22) \citet{str99};
(23) \citet{tin03}}
\end{deluxetable}

\begin{deluxetable}{lrccccccc}
\tabletypesize{\scriptsize}
\tablecaption{Number Densities as a Function of Systemic Spectral Type\label{tab_sptdist}}
\tablewidth{0pt}
\tablehead{
\colhead{} &
\multicolumn{1}{r|}{$\Psi$(M) $\propto$} &
\colhead{M$^{0.0}$} &
\colhead{M$^{-0.5}$} &
\colhead{M$^{-1.0}$} &
\colhead{M$^{-1.5}$} &
\colhead{lognormal} &
\colhead{M$^{-0.5}$} &
\colhead{M$^{-0.5}$} \\
\colhead{} &
\multicolumn{1}{r|}{$P(q) \propto$} &
\colhead{$q^4$} &
\colhead{$q^4$} &
\colhead{$q^4$} &
\colhead{$q^4$} &
\colhead{$q^4$} &
\colhead{$q^4$} &
\colhead{$q^4$} \\
\colhead{} &
\multicolumn{1}{r|}{${\epsilon}_b$ =} &
\colhead{0.1} &
\colhead{0.1} &
\colhead{0.1} &
\colhead{0.1} &
\colhead{0.1} &
\colhead{0.3} &
\colhead{0.5} \\
}
\startdata
L0--L1 &  & 0.555 & 0.637 & 0.712 & 0.812 & 0.692 & 0.503 & 0.396 \\
L1--L2 &  & 0.755 & 0.861 & 1.01 & 1.16 & 0.973 & 0.869 & 0.905 \\
L2--L3 &  & 0.848 & 1.00 & 1.16 & 1.37 & 1.12 & 1.01 & 1.01 \\
L3--L4 &  & 0.693 & 0.834 & 1.01 & 1.25 & 0.962 & 0.787 & 0.756 \\
L4--L5 &  & 0.672 & 0.819 & 1.03 & 1.33 & 0.964 & 0.767 & 0.707 \\
L5--L6 &  & 0.688 & 0.869 & 1.14 & 1.52 & 1.03 & 0.911 & 0.945 \\
L6--L7 &  & 0.618 & 0.785 & 1.04 & 1.39 & 0.914 & 0.764 & 0.746 \\
L7--L8 &  & 0.502 & 0.644 & 0.858 & 1.18 & 0.772 & 0.614 & 0.579 \\
L8--L9 &  & 0.351 & 0.462 & 0.607 & 0.856 & 0.555 & 0.428 & 0.393 \\
L9--T0 &  & 0.237 & 0.308 & 0.421 & 0.597 & 0.366 & 0.328 & 0.342 \\
T0--T1 &  & 0.165 & 0.220 & 0.294 & 0.412 & 0.261 & 0.251 & 0.297 \\
T1--T2 &  & 0.181 & 0.241 & 0.326 & 0.442 & 0.284 & 0.300 & 0.362 \\
T2--T3 &  & 0.304 & 0.406 & 0.549 & 0.780 & 0.475 & 0.452 & 0.503 \\
T3--T4 &  & 0.605 & 0.788 & 1.08 & 1.54 & 0.949 & 0.845 & 0.931 \\
T4--T5 &  & 1.06 & 1.42 & 1.94 & 2.76 & 1.66 & 1.40 & 1.41 \\
T5--T6 &  & 1.71 & 2.38 & 3.40 & 5.06 & 2.88 & 2.60 & 2.83 \\
T6--T7 &  & 2.39 & 3.45 & 5.09 & 7.79 & 4.18 & 3.23 & 3.01 \\
T7--T8 &  & 3.14 & 4.82 & 7.63 & 12.4 & 5.94 & 4.82 & 4.86 \\
\enddata
\tablecomments{Number densities in units of 10$^{-3}$ pc$^{-3}$ SpT$^{-1}$.
Results listed here are based on simulations using the
evolutionary models of \citet{bar03}, a flat age distribution spanning
0.01--10~Gyr, the $M_K$/spectral type relation defined in this study ($\S$~3.5.1),
and parameters as specified in the header.  Values
are normalized to a number density of 0.0037 pc$^{-3}$ over the
mass range of 0.09--0.1~M$_{\sun}$, derived from the low-mass stellar mass
function of \citet{rei99}.}
\end{deluxetable}

\begin{deluxetable}{lrcccccccccccc}
\tabletypesize{\scriptsize}
\rotate
\tablecaption{Surface Densities for a Magnitude-Limited Sample as a Function of Systemic Spectral Type\label{tab_sptdistmag}}
\tablewidth{0pt}
\tablehead{
\colhead{} &
\multicolumn{1}{r|}{$\Psi$(M) $\propto$} &
\colhead{M$^{0.0}$} &
\colhead{M$^{-0.5}$} &
\colhead{M$^{-1.0}$} &
\colhead{M$^{-1.5}$} &
\colhead{lognormal} &
\colhead{M$^{-0.5}$} &
\colhead{M$^{-0.5}$} &
\colhead{M$^{-0.5}$} &
\colhead{M$^{-0.5}$} &
\colhead{M$^{-0.5}$} &
\colhead{M$^{-0.5}$} &
\colhead{M$^{-0.5}$} \\
\colhead{} &
\multicolumn{1}{r|}{$M$(SpT):\tablenotemark{a}} &
\colhead{TP-$K$} &
\colhead{TP-$K$} &
\colhead{TP-$K$} &
\colhead{TP-$K$} &
\colhead{TP-$K$} &
\colhead{L06-$K$} &
\colhead{L06p-$K$} &
\colhead{L06-$J$} &
\colhead{L06p-$J$} &
\colhead{TP-$K$} &
\colhead{TP-$K$} &
\colhead{TP-$K$} \\
\colhead{SpT} &
\multicolumn{1}{r|}{$P(q) \propto$} &
\colhead{$q^4$} &
\colhead{$q^4$} &
\colhead{$q^4$} &
\colhead{$q^4$} &
\colhead{$q^4$} &
\colhead{$q^4$} &
\colhead{$q^4$} &
\colhead{$q^4$} &
\colhead{$q^4$} &
\colhead{1} &
\colhead{$q^4$} &
\colhead{$q^4$} \\
\colhead{} &
\multicolumn{1}{r|}{${\epsilon}_b$ =} &
\colhead{0.1} &
\colhead{0.1} &
\colhead{0.1} &
\colhead{0.1} &
\colhead{0.1} &
\colhead{0.1} &
\colhead{0.1} &
\colhead{0.1} &
\colhead{0.1} &
\colhead{0.1} &
\colhead{0.3} &
\colhead{0.5} \\
}
\startdata
L0--L1 &  & 8.24 & 9.52 & 10.6 & 12.1 & 10.3 & 8.62 & 12.4 & 1.87 & 2.49 & 9.56 & 7.92 & 6.69 \\
L1--L2 &  & 8.68 & 9.96 & 11.6 & 13.5 & 11.2 & 11.5 & 11.0 & 2.00 & 1.94 & 8.96 & 12.2 & 15.0 \\
L2--L3 &  & 6.58 & 7.80 & 8.99 & 10.7 & 8.70 & 8.02 & 6.85 & 1.13 & 1.02 & 7.41 & 9.02 & 10.4 \\
L3--L4 &  & 3.63 & 4.36 & 5.28 & 6.46 & 5.02 & 3.77 & 3.41 & 0.430 & 0.408 & 4.27 & 4.76 & 5.33 \\
L4--L5 &  & 2.21 & 2.68 & 3.37 & 4.30 & 3.13 & 2.02 & 2.06 & 0.205 & 0.195 & 2.63 & 3.00 & 3.29 \\
L5--L6 &  & 1.47 & 1.84 & 2.40 & 3.19 & 2.19 & 1.38 & 1.48 & 0.125 & 0.126 & 1.62 & 2.56 & 3.28 \\
L6--L7 &  & 0.736 & 0.932 & 1.22 & 1.63 & 1.08 & 0.818 & 0.806 & 0.0697 & 0.0678 & 0.885 & 1.14 & 1.34 \\
L7--L8 &  & 0.398 & 0.500 & 0.663 & 0.894 & 0.600 & 0.518 & 0.449 & 0.0497 & 0.0424 & 0.494 & 0.639 & 0.761 \\
L8--L9 &  & 0.194 & 0.254 & 0.325 & 0.464 & 0.305 & 0.327 & 0.239 & 0.0371 & 0.0261 & 0.245 & 0.316 & 0.358 \\
L9--T0 &  & 0.119 & 0.151 & 0.205 & 0.285 & 0.182 & 0.227 & 0.130 & 0.0328 & 0.0201 & 0.147 & 0.242 & 0.320 \\
T0--T1 &  & 0.0761 & 0.100 & 0.131 & 0.186 & 0.118 & 0.164 & 0.0777 & 0.0287 & 0.0163 & 0.0882 & 0.171 & 0.242 \\
T1--T2 &  & 0.0656 & 0.0869 & 0.117 & 0.160 & 0.102 & 0.156 & 0.0623 & 0.0441 & 0.0199 & 0.0735 & 0.150 & 0.214 \\
T2--T3 &  & 0.0712 & 0.0952 & 0.129 & 0.179 & 0.110 & 0.171 & 0.0624 & 0.0738 & 0.0316 & 0.0857 & 0.143 & 0.190 \\
T3--T4 &  & 0.0781 & 0.102 & 0.138 & 0.199 & 0.123 & 0.179 & 0.0706 & 0.131 & 0.0611 & 0.0935 & 0.134 & 0.171 \\
T4--T5 &  & 0.0670 & 0.0897 & 0.122 & 0.173 & 0.104 & 0.142 & 0.0749 & 0.164 & 0.0948 & 0.0855 & 0.102 & 0.117 \\
T5--T6 &  & 0.0510 & 0.0711 & 0.101 & 0.151 & 0.0859 & 0.0968 & 0.0705 & 0.161 & 0.121 & 0.0644 & 0.0910 & 0.111 \\
T6--T7 &  & 0.0307 & 0.0443 & 0.0652 & 0.0996 & 0.0536 & 0.0483 & 0.0463 & 0.0975 & 0.0926 & 0.0441 & 0.0462 & 0.0481 \\
T7--T8 &  & 0.0184 & 0.0281 & 0.0442 & 0.0713 & 0.0346 & 0.0260 & 0.0282 & 0.0527 & 0.0565 & 0.0265 & 0.0337 & 0.0397 \\
\enddata
\tablecomments{Surface densities in 10$^{-3}$ deg$^{-2}$ SpT$^{-1}$.
Results listed here are based on simulations using the
evolutionary models of \citet{bar03}, a flat age distribution spanning
0.01--10~Gyr, and parameters as specified in the header.  Values
are normalized to a number density of 0.0037 pc$^{-3}$ over the
mass range of 0.09--0.1~M$_{\sun}$, derived from the low-mass stellar mass
function of \citet{rei99}; and assume a limiting magnitude of 16 in the 
appropriate photometric band.}
\tablenotetext{a}{Absolute magnitude/spectral type relations
from (TP): this paper; (L06): \citet{liu06}, excluding known binaries;
(L06p): \citet{liu06}, excluding known and possible binaries 
(see $\S$~3.5.1).}
\end{deluxetable}

\begin{deluxetable}{lrcccccccccc}
\tabletypesize{\scriptsize}
\tablecaption{Binary Fraction as a Function of Systemic Spectral Type: Volume-Limited\label{tab_bfdist}}
\tablewidth{0pt}
\tablehead{
\colhead{} &
\multicolumn{1}{r|}{$\Psi$(M) $\propto$} &
\colhead{M$^{-0.5}$} &
\colhead{M$^{-0.5}$} &
\colhead{M$^{-0.5}$} &
\colhead{M$^{-0.5}$} &
\colhead{M$^{-0.5}$} &
\colhead{M$^{-0.5}$} &
\colhead{M$^{-0.5}$} &
\colhead{M$^{-0.5}$} &
\colhead{M$^{-0.5}$} &
\colhead{M$^{-0.5}$} \\
\colhead{} &
\multicolumn{1}{r|}{$M$(SpT):\tablenotemark{a}} &
\colhead{TP-$K$} &
\colhead{TP-$K$} &
\colhead{TP-$K$} &
\colhead{TP-$K$} &
\colhead{TP-$K$} &
\colhead{L06-$K$} &
\colhead{L06p-$K$} &
\colhead{L06-$J$} &
\colhead{L06p-$J$} &
\colhead{TP-$K$} \\
\colhead{SpT} &
\multicolumn{1}{r|}{$P(q) \propto$} &
\colhead{$q^4$} &
\colhead{$q^4$} &
\colhead{$q^4$} &
\colhead{$q^4$} &
\colhead{$q^4$} &
\colhead{$q^4$} &
\colhead{$q^4$} &
\colhead{$q^4$} &
\colhead{$q^4$} &
\colhead{1} \\
\colhead{} &
\multicolumn{1}{r|}{${\epsilon}_b$ =} &
\colhead{0.1} &
\colhead{0.2} &
\colhead{0.3} &
\colhead{0.4} &
\colhead{0.5} &
\colhead{0.1} &
\colhead{0.1} &
\colhead{0.1} &
\colhead{0.1} &
\colhead{0.1} \\
}
\startdata
L0--L1 &  & 0.0168 & 0.0344 & 0.0635 & 0.0888 & 0.127 & 0.0122 & 0.0239 & 0.0178 & 0.0272 & 0.0455 \\
L1--L2 &  & 0.111 & 0.218 & 0.318 & 0.424 & 0.526 & 0.109 & 0.116 & 0.113 & 0.117 & 0.0967 \\
L2--L3 &  & 0.108 & 0.215 & 0.311 & 0.408 & 0.522 & 0.105 & 0.101 & 0.104 & 0.104 & 0.109 \\
L3--L4 &  & 0.0788 & 0.162 & 0.247 & 0.333 & 0.435 & 0.0758 & 0.0726 & 0.0759 & 0.0780 & 0.0857 \\
L4--L5 &  & 0.0678 & 0.137 & 0.219 & 0.304 & 0.391 & 0.0627 & 0.0648 & 0.0598 & 0.0574 & 0.0782 \\
L5--L6 &  & 0.119 & 0.229 & 0.339 & 0.446 & 0.547 & 0.110 & 0.114 & 0.106 & 0.112 & 0.0910 \\
L6--L7 &  & 0.0911 & 0.178 & 0.275 & 0.373 & 0.467 & 0.0813 & 0.0919 & 0.0719 & 0.0763 & 0.0932 \\
L7--L8 &  & 0.0755 & 0.154 & 0.245 & 0.331 & 0.422 & 0.0667 & 0.0799 & 0.0647 & 0.0717 & 0.100 \\
L8--L9 &  & 0.0666 & 0.146 & 0.226 & 0.309 & 0.391 & 0.0627 & 0.0775 & 0.0649 & 0.0653 & 0.0873 \\
L9--T0 &  & 0.114 & 0.228 & 0.357 & 0.455 & 0.563 & 0.115 & 0.123 & 0.107 & 0.119 & 0.128 \\
T0--T1 &  & 0.180 & 0.326 & 0.459 & 0.573 & 0.651 & 0.204 & 0.176 & 0.184 & 0.192 & 0.168 \\
T1--T2 &  & 0.217 & 0.393 & 0.521 & 0.643 & 0.720 & 0.272 & 0.208 & 0.281 & 0.249 & 0.186 \\
T2--T3 &  & 0.159 & 0.294 & 0.416 & 0.521 & 0.621 & 0.187 & 0.141 & 0.200 & 0.162 & 0.138 \\
T3--T4 &  & 0.139 & 0.255 & 0.379 & 0.489 & 0.587 & 0.141 & 0.125 & 0.147 & 0.133 & 0.113 \\
T4--T5 &  & 0.0997 & 0.196 & 0.297 & 0.398 & 0.497 & 0.0983 & 0.106 & 0.0955 & 0.101 & 0.104 \\
T5--T6 &  & 0.142 & 0.273 & 0.394 & 0.502 & 0.597 & 0.141 & 0.141 & 0.141 & 0.142 & 0.112 \\
T6--T7 &  & 0.0718 & 0.147 & 0.228 & 0.316 & 0.409 & 0.0720 & 0.0719 & 0.0727 & 0.0727 & 0.0931 \\
T7--T8 &  & 0.0975 & 0.201 & 0.298 & 0.399 & 0.501 & 0.0995 & 0.102 & 0.101 & 0.101 & 0.101 \\
\enddata
\tablecomments{Results based on simulations using the
evolutionary models of \citet{bar03}, a flat age distribution spanning
0.01--10~Gyr, and parameters as indicated in the header.}
\tablenotetext{a}{Absolute magnitude/spectral type relations
from (TP): this paper; (L06): \citet{liu06}, excluding known binaries;
(L06p): \citet{liu06}, excluding known and possible binaries 
(see $\S$~3.5.1).}
\end{deluxetable}

\begin{deluxetable}{lrcccccccccc}
\tabletypesize{\scriptsize}
\tablecaption{Binary Fraction as a Function of Systemic Spectral Type: Magnitude-Limited\label{tab_bfdistmag}}
\tablewidth{0pt}
\tablehead{
\colhead{} &
\multicolumn{1}{r|}{$\Psi$(M) $\propto$} &
\colhead{M$^{-0.5}$} &
\colhead{M$^{-0.5}$} &
\colhead{M$^{-0.5}$} &
\colhead{M$^{-0.5}$} &
\colhead{M$^{-0.5}$} &
\colhead{M$^{-0.5}$} &
\colhead{M$^{-0.5}$} &
\colhead{M$^{-0.5}$} &
\colhead{M$^{-0.5}$} &
\colhead{M$^{-0.5}$} \\
\colhead{} &
\multicolumn{1}{r|}{$M$(SpT):\tablenotemark{a}} &
\colhead{TP-$K$} &
\colhead{TP-$K$} &
\colhead{TP-$K$} &
\colhead{TP-$K$} &
\colhead{TP-$K$} &
\colhead{L06-$K$} &
\colhead{L06p-$K$} &
\colhead{L06-$J$} &
\colhead{L06p-$J$} &
\colhead{TP-$K$} \\
\colhead{SpT} &
\multicolumn{1}{r|}{$P(q) \propto$} &
\colhead{$q^4$} &
\colhead{$q^4$} &
\colhead{$q^4$} &
\colhead{$q^4$} &
\colhead{$q^4$} &
\colhead{$q^4$} &
\colhead{$q^4$} &
\colhead{$q^4$} &
\colhead{$q^4$} &
\colhead{1} \\
\colhead{} &
\multicolumn{1}{r|}{${\epsilon}_b$ =} &
\colhead{0.1} &
\colhead{0.2} &
\colhead{0.3} &
\colhead{0.4} &
\colhead{0.5} &
\colhead{0.1} &
\colhead{0.1} &
\colhead{0.1} &
\colhead{0.1} &
\colhead{0.1} \\
}
\startdata
L0--L1 &  & 0.0370 & 0.0737 & 0.131 & 0.181 & 0.244 & 0.0267 & 0.0461 & 0.0394 & 0.0507 & 0.0498 \\
L1--L2 &  & 0.214 & 0.378 & 0.506 & 0.615 & 0.707 & 0.193 & 0.220 & 0.198 & 0.223 & 0.135 \\
L2--L3 &  & 0.179 & 0.331 & 0.449 & 0.556 & 0.665 & 0.172 & 0.171 & 0.184 & 0.188 & 0.145 \\
L3--L4 &  & 0.148 & 0.286 & 0.399 & 0.506 & 0.614 & 0.158 & 0.147 & 0.184 & 0.179 & 0.122 \\
L4--L5 &  & 0.150 & 0.278 & 0.406 & 0.515 & 0.610 & 0.169 & 0.150 & 0.229 & 0.166 & 0.121 \\
L5--L6 &  & 0.268 & 0.453 & 0.586 & 0.693 & 0.771 & 0.283 & 0.262 & 0.311 & 0.293 & 0.162 \\
L6--L7 &  & 0.204 & 0.359 & 0.496 & 0.604 & 0.692 & 0.215 & 0.204 & 0.250 & 0.220 & 0.169 \\
L7--L8 &  & 0.214 & 0.381 & 0.521 & 0.622 & 0.710 & 0.180 & 0.197 & 0.238 & 0.249 & 0.212 \\
L8--L9 &  & 0.200 & 0.372 & 0.506 & 0.609 & 0.685 & 0.167 & 0.207 & 0.225 & 0.224 & 0.183 \\
L9--T0 &  & 0.317 & 0.530 & 0.671 & 0.755 & 0.824 & 0.264 & 0.327 & 0.314 & 0.362 & 0.300 \\
T0--T1 &  & 0.428 & 0.621 & 0.746 & 0.820 & 0.863 & 0.387 & 0.472 & 0.354 & 0.452 & 0.369 \\
T1--T2 &  & 0.435 & 0.640 & 0.751 & 0.831 & 0.877 & 0.453 & 0.499 & 0.421 & 0.466 & 0.356 \\
T2--T3 &  & 0.330 & 0.525 & 0.654 & 0.741 & 0.812 & 0.351 & 0.365 & 0.281 & 0.294 & 0.255 \\
T3--T4 &  & 0.249 & 0.413 & 0.558 & 0.662 & 0.747 & 0.242 & 0.256 & 0.248 & 0.257 & 0.173 \\
T4--T5 &  & 0.174 & 0.316 & 0.441 & 0.556 & 0.649 & 0.158 & 0.191 & 0.161 & 0.187 & 0.142 \\
T5--T6 &  & 0.228 & 0.397 & 0.535 & 0.640 & 0.723 & 0.206 & 0.231 & 0.220 & 0.243 & 0.149 \\
T6--T7 &  & 0.119 & 0.233 & 0.343 & 0.448 & 0.550 & 0.112 & 0.119 & 0.116 & 0.122 & 0.119 \\
T7--T8 &  & 0.185 & 0.346 & 0.472 & 0.582 & 0.678 & 0.189 & 0.194 & 0.189 & 0.189 & 0.140 \\
\enddata
\tablecomments{Simulations based on the
evolutionary models of \citet{bar03}, a flat age distribution spanning
0.01--10~Gyr, and parameters as indicated in the header.  A limiting magnitude
of 16 in the respective filter band is assumed.}
\tablenotetext{a}{Absolute magnitude/spectral type relations
from (TP): this paper; (L06): \citet{liu06}, excluding known binaries;
(L06p): \citet{liu06}, excluding known and possible binaries 
(see $\S$~3.5.1).}
\end{deluxetable}

\clearpage

\end{document}